\documentclass[11pt,peerreviewca,draftclsnofoot]{IEEEtran}
\usepackage{eurosym}
\usepackage{cite,algorithm,array,url}
\usepackage{graphicx,colortbl}
\usepackage{amsfonts,times,amsmath,amssymb,amsthm}

\theoremstyle{plain}
\newtheorem{theorem}{Theorem}
\newtheorem{lemma}[theorem]{Lemma}
\newtheorem{proposition}[theorem]{Proposition}

\theoremstyle{definition}

\theoremstyle{remark}

\input{tcilatex}
\begin{document}

\title{Multi-Sensor Multi-object Tracking with the Generalized Labeled
Multi-Bernoulli Filter}
\author{Ba-Ngu~Vo and Ba-Tuong~Vo  \thanks{%
B.-N. Vo and B.-T. Vo are with the Department of Electrical and Computer
Engineering, Curtin University, Bentley, WA 6102, Australia (email:
\{ba-tuong.vo,ba-ngu.vo\}@curtin.edu.au). }}
\maketitle

\begin{abstract}
This paper proposes an efficient implementation of the multi-sensor
generalized labeled multi-Bernoulli (GLMB) filter. The solution exploits the
GLMB joint prediction and update together with a new technique for
truncating the GLMB filtering density based on Gibbs sampling. The resulting
algorithm has quadratic complexity in the number of hypothesized object and
linear in the number of measurements of each individual sensors.
\end{abstract}

%\markboth{Preprint: IEEE Transactions on Signal Processing,~Vol.~XX, No.~X,
%pp. X--X, ~XXX~2014}{Hoang \MakeLowercase{\textit{et al.}}: The Cauchy-Schwarz divergence for Poisson point processes}





\section{Introduction}

The objective of multi-object tracking is to jointly estimate the number of
objects and their trajectories from sensor data \cite{BSF88,BP99, Mah07,
mahler2014advances}. A majority of multi-object tracking techniques are
developed for single sensors. The use of multiple sensors, in principle,
reduces uncertainty about the object existence as well as its states.
However, this problem is computationally intractable in general, especially
for more than two sensors, even though conceptually the generalization to
multiple sensors can be straightforward.

The random finite set (RFS) framework developed by Mahler \cite{Mah07,
mahler2014advances} has attracted significant attention as a general
systematic treatment of multi-sensor multi-object systems. This framework
facilitates the development of novel filters such as the Probability
Hypothesis Density (PHD) filter \cite{MahlerPHD}, Cardinalized PHD (CPHD)
filter \cite{MahlerCPHD}, and multi-Bernoulli filters \cite%
{Mah07,VVC09,VVPS10}. While these filters were not designed to estimate the
trajectories of objects, they have been successfully deployed in many
applications including radar/sonar \cite{TobiasLanterman05}, \cite%
{ClarkBell05}, computer vision \cite{MTC_CSVT08,HVVS_PR12,HVV_TSP13}, cell
biology \cite{Rez_TMI15}, autonomous vehicle \cite{MVA_TRO11, LHG_TSP11,
LCS_SLAM_STSP13} automotive safety \cite{Bat08,MRD13}, sensor scheduling 
\cite{RVC11,HV14sencon, Gostar13, HVVM15, Gostar16, GostarSP16, Gostar17},
and sensor network \cite{Zhang_TAC11,BCF_STSP13,UCJ_STSP13}.

The classical PHD and CPHD filters are developed for single-sensors. Since
the multi-sensor PHD, CPHD and multi-Bernoulli filters are combinatiorial 
\cite{mahler2014advances}, \cite{Saucanetal16}, the most commonly used
approximate multi-sensor PHD, CPHD and multi-Bernoulli filter are the
heuristic \textquotedblleft iterated corrector\textquotedblright\ versions 
\cite{VSM04} that apply single-sensor updates, once for each sensor in turn.
This approach yields final solutions that depend on the order in which the
sensors are processed. Multi-sensor PHD and CPHD filters that are
principled, computationally tractable, and independent of sensor order have
been proposed in \cite{mahler2014advances} (Section 10.6). However, this
approach as well as the heuristic \textquotedblleft iterated
corrector\textquotedblright\ involve two levels of approximation since the
exact multi-sensor PHD, CPHD and multi-Bernoulli filters are approximations
of the Bayes multi-sensor multi-object filter.

An exact solution to the Bayes multi-object filter is the Generalized
Labeled Multi-Bernoulli (GLMB) filter, which also outputs multi-object
trajectories \cite{VoGLMB13}, \cite{VVP_GLMB13}. Moreover, given a cap on
the number of GLMB components, recent works show that the GLMB filter can be
implemented with linear complexity in the number of measurements and
quadratic in the number of hypothesized objects \cite{VVH16}. The GLMB
density is flexible enough to approximate any labeled RFS density with
matching intensity function and cardinality distribution \cite{Papi_etal15},
and also enjoys a number of nice analytical properties, e.g. the void
probability functional--a necessary and sufficient statistic--of a GLMB, the
Cauchy-Schwarz divergence between two GLMBs, the $L_{1}$-distance between a
GLMB and its truncation, can all be computed in closed form \cite{BVVA15}, 
\cite{VVP_GLMB13}. Recent research in approximate GLMB filters \cite%
{ReuterLMB14, Fantacci_etal16} as well as applications in tracking from
merged measurements \cite{BVV15}, extended targets \cite{Beard_etal16},
maneuvering targets \cite{PVV16, RezaFusion16}, track-before-detect \cite%
{PapiKim15, Papi_etal15}, computer vision \cite{Kim13, Punchihewa14,
Rathnayake15, KVV16}, sensor scheduling \cite{Gostar14, BVVA15}, field
robotics \cite{Deusch15}, and distributed multi-object tracking \cite%
{Fantacci_etal15}, demonstrate the versatility of the GLMB filter, and
suggest that it is an important tool in multi-object systems.

In this work we present an implementation of the multi-sensor GLMB filter.
The major hurdle in the multi-sensor GLMB filter implementation is the
NP-hard multi-dimensional ranked assignment problem. A multi-sensor version
of an approximation of the GLMB filter, known as the marginalized GLMB
filter, was proposed in \cite{Fantacci_etal16}. While this multi-sensor
solution is scalable in the number of sensors, it still involves two levels
of approximations: the truncation of the GLMB density; and the functional
approximation of the truncated GLMB density. An implementation of the
two-sensor GLMB filter was developed in \cite{Wei16} using Murty's
algorithm. This implementation has a cubic complexity in the product of the
number of measurements from the sensors. The \textquotedblleft iterated
corrector\textquotedblright\ strategy would yield the exact solution if all
the GLMB components are kept. However in practice truncation is performed at
each single-sensor update, which leads to a final solution that depends on
the order of the sensor updates. More importantly, an extremely large number
of GLMB components would be needed in the process even if the final GLMB
filtering density only contains a small number of components. Components
that are significant after one single-sensor update may not be significant
at another update. Worse, insignificant components after one single sensor
update, which could become significant in the final GLMB filtering density,
are discarded and cannot be recovered. To circumvent these problems, we
extend the GLMB truncation technique based on Gibbs sampling proposed in 
\cite{VVH16} to the multi-sensor case.

\section{Background}

This section summarises the multi-object state space models and the GLMB
filter.

\subsection{Multi-object State}

\indent At time $k$, an existing object is described by a vector $x_{k}\in 
\mathbb{X}$. To distinguish different object trajectories, each object is
assigned a unique label $\ell _{k}$ that consists of an ordered pair $(t,i)$%
, where $t$ is the time of birth and $i$ is the index of individual objects
born at the same time \cite{VoGLMB13}. The trajectory or track of an object
is given by the sequence of states with the same label.

Formally, the state of an object at time $k$ is a vector $\mathbf{x}%
_{k}=(x_{k},\ell _{k})\in \mathbb{X\times L}_{k}$, where $\mathbb{L}_{k}$
denotes the label space for objects at time $k$ (including those born prior
to $k$). Note that $\mathbb{L}_{k}$ is given by $\mathbb{B}_{k}\cup \mathbb{L%
}_{k-1}$, where $\mathbb{B}_{k}$ denotes the label space for objects born at
time $k$ (and is disjoint from $\mathbb{L}_{k-1}$). Suppose that there are $%
N_{k}$ objects at time $k$, with states $\mathbf{x}_{k,1},...,\mathbf{x}%
_{k,N_{k}}$, in the context of multi-object tracking, the collection of
states, referred to as the \emph{multi-object state}, is naturally
represented as a finite set 
\begin{equation*}
\mathbf{X}_{k}=\{\mathbf{x}_{k,1},...,\mathbf{x}_{k,N_{k}}\}\in \mathcal{F}(%
\mathbb{X\times L}_{k}),
\end{equation*}%
where $\mathcal{F}(\mathbb{X\times L}_{k})$ denotes the space of finite
subsets of $\mathbb{X\times L}_{k}$. We denote cardinality (number of
elements) of $\mathbf{X}$ by $|\mathbf{X}|$ and the set of labels of $%
\mathbf{X}$, $\{\ell :(x,\ell )\in \mathbf{X}\}$, by $\mathcal{L}_{\mathbf{X}%
}$. Note that since the label is unique, no two objects have the same label,
i.e. $\delta _{|\mathbf{X}|}(|\mathcal{L}_{\mathbf{X}}|)=1$. Hence $\Delta (%
\mathbf{X})\triangleq $ $\delta _{|\mathbf{X}|}(|\mathcal{L}_{\mathbf{X}}|)$
is called the \emph{distinct label indicator}.

\indent For the rest of the paper, we follow the convention that
single-object states are represented by lower-case letters (e.g. $x$, $%
\mathbf{x}$), while multi-object states are represented by upper-case
letters (e.g. $X$, $\mathbf{X}$), symbols for labeled states and their
distributions are bold-faced to distinguish them from unlabeled ones (e.g. $%
\mathbf{x}$, $\mathbf{X}$, $\mathbf{\pi }$, etc.), spaces are represented by
blackboard bold (e.g. $\mathbb{X}$, $\mathbb{Z}$, $\mathbb{L}$, $\mathbb{N}$%
, etc.). The inner product $\int f(x)g(x)dx$ is denoted by $\left\langle
f,g\right\rangle $. The list of variables $X_{m},X_{m+1},...,X_{n}$ is
abbreviated as $X_{m:n}$. For a finite set $X$, its cardinality (or number
of elements) is denoted by $|X|$, in addition we use the multi-object
exponential notation $f^{X}$ for the product $\tprod_{x\in X}f(x)$, with $%
f^{\emptyset }=1$. We denote a generalization of the Kroneker delta that
takes arbitrary arguments such as sets, vectors, integers etc., by 
\begin{equation*}
\delta _{Y}[X]\triangleq \left\{ 
\begin{array}{l}
1,\text{ if }X=Y \\ 
0,\text{ otherwise}%
\end{array}%
\right. .
\end{equation*}%
For a given set $S$, $1_{S}(\cdot )$ denotes the indicator function of $S$,
and $\mathcal{F}(S)$ denotes the class of finite subsets of $S$. Also, for
notational compactness, we drop the subscript $k$ for the current time, the
next time is indicated by the subscript `$+$'.

\subsection{Standard multi-object dynamic model}

Given the multi-object state $\mathbf{X}$ (at time $k$), each state $(x,\ell
)\in \mathbf{X}$ either survives with probability $P_{S}(x,\ell )$ and
evolves to a new state $(x_{+},\ell _{+})$ (at time $k+1$) with probability
density $f_{+}(x_{+}|x,\ell )\delta _{\ell }[\ell _{+}]$ or dies with
probability $1-P_{S}(x,\ell )$. The set $\mathbf{B}_{+}$ of new objects
(born at time $k+1)$ is distributed according to the labeled multi-Bernoulli
(LMB) density\footnote{%
Note that in this work we use Mahler's set derivatives for multi-object
densities \cite{Mah07, mahler2014advances}. While these are not actual
probability densities, they are equivalent to probability densities relative
to a certain reference measure \cite{VSD05}.} 
\begin{equation}
\mathbf{f}_{B,+}(\mathbf{B}_{+})=\Delta (\mathbf{B}_{+})\left[ 1_{\mathbb{B}%
_{\,+}}\,r_{B,+}\right] ^{\mathcal{L(}\mathbf{B}_{+})}\left[ 1-r_{B,+}\right]
^{\mathbb{B}_{+}-\mathcal{L(}\mathbf{B}_{+})}p_{B,+}^{\mathbf{B}_{+}},
\label{eq:LMB_birth}
\end{equation}%
where $r_{B,+}(\ell )$ is the probability that a new object with label $\ell 
$ is born, and $p_{B,+}(\cdot ,\ell )$\ is the distribution of its kinematic
state \cite{VoGLMB13}. The multi-object state $\mathbf{X}_{+}$ (at time $k+1$%
) is the superposition of surviving objects and new born objects. It is
assumed that, conditional on $\mathbf{X}$, \ objects move, appear and die
independently of each other. The expression for the multi-object transition
density $\mathbf{f}_{+}$ is given by \cite{VoGLMB13}, \cite{VVP_GLMB13} 
\begin{equation}
\mathbf{f}_{+}\left( \mathbf{X}_{+}|\mathbf{X}\right) =\mathbf{f}_{S,+}(%
\mathbf{X}_{+}\cap (\mathbb{X}\times \mathbb{L)}|\mathbf{X})\mathbf{f}_{B,+}(%
\mathbf{X}_{+}-(\mathbb{X}\times \mathbb{L}))  \label{eq:labeled_transition}
\end{equation}%
where 
\begin{eqnarray}
\!\!\mathbf{f}_{S,+}(\mathbf{W}|\mathbf{X})\!\!\! &=&\!\!\!\Delta (\mathbf{W}%
)\Delta (\mathbf{X})1_{\mathcal{L}(\mathbf{X})}(\mathcal{L(}\mathbf{W}))%
\left[ \Phi (\mathbf{W};\cdot )\right] ^{\mathbf{X}}
\label{eq:Survival_transition} \\
\!\!\Phi (\mathbf{W};x,\ell )\!\!\! &=&\!\!\!\left\{ \!\!%
\begin{array}{ll}
P_{S}(x,\ell )f_{+}(x_{+}|x,\ell ), & \!\!\text{if }\left( x_{+},\ell
\right) \in \mathbf{W} \\ 
1-P_{S}(x,\ell ), & \!\!\text{if }\ell \notin \mathcal{L}(\mathbf{W})%
\end{array}%
\right. \!\!.  \label{eq:Survival_transition2}
\end{eqnarray}

\subsection{Standard multi-object observation model}

For a given multi-object state $\mathbf{X}$, each $(x,\ell )\in \mathbf{X}$
is either detected by sensor $s$ with probability $P_{D}^{(s)}(x,\ell )$ and
generates a detection $z^{(s)}\in Z^{(s)}$ with likelihood $%
g_{D}^{(s)}(z^{(s)}|x,\ell )$ or missed with probability $%
1-P_{D}^{(s)}(x,\ell )$. The \emph{multi-object observation} is the
superposition of the observations from detected objects and Poisson clutter
with intensity $\kappa ^{(s)}$.

Assuming that, conditional on $\mathbf{X}$, detections are independent of
each other and clutter, the multi-object likelihood function of sensor $s$
is given by \cite{VoGLMB13}, \cite{VVP_GLMB13} 
\begin{equation}
g^{(s)}(Z^{(s)}|\mathbf{X})\propto \sum_{\theta ^{(s)}\in \Theta
^{(s)}}1_{\Theta ^{(s)}(\mathcal{L(}\mathbf{X}))}(\theta
^{(s)})\prod\limits_{(x,\ell )\in \mathbf{X}}\psi
_{Z_{_{\!}}^{(s)}}^{(s,\theta ^{(s)}(\ell ))}(x,\ell )
\label{eq:RFSmeaslikelihood0}
\end{equation}%
where: $\Theta ^{(s)}$ is the set of \emph{positive 1-1} maps $\theta ^{(s)}:%
\mathbb{L}\rightarrow \{0$:$|Z^{(s)}|\}$, i.e. maps such that \emph{no two
distinct arguments are mapped to the same positive value}, $\Theta ^{(s)}(I)$
is the subset of $\Theta ^{(s)}$ with domain $I$; and 
\begin{equation}
\psi _{\!\{z_{1:M^{(s)}}\}\!}^{(s,j)}(x,\ell )=\left\{ 
\begin{array}{ll}
\frac{P_{\!D}^{(s)}(x,\ell )g^{(s)}(z_{j}|x,\ell )}{\kappa ^{(s)}(z_{j})}, & 
\!\!\text{if }j=1\text{:}M^{(s)} \\ 
1-P_{\!D}^{(s)}(x,\ell ), & \!\!\text{if }j=0%
\end{array}%
\right. .  \label{eq:PropConj5}
\end{equation}%
The map $\theta ^{(s)}$ specifies which objects generated which detections
from sensor $s$, i.e. object $\ell $ generates detection $z_{\theta (\ell
)}\in Z^{(s)}$, with undetected objects assigned to $0$. The positive 1-1
property means that $\theta ^{(s)}$ is 1-1 on $\{\ell :\theta ^{(s)}(\ell
)>0\}$, the set of labels that are assigned positive values, and ensures
that any detection in $Z^{(s)}$ is assigned to at most one object.

Assuming that the sensors are conditionally independent, the multi-sensor
likelihood \ is given by%
\begin{eqnarray}
g(Z^{(1)},...,Z^{(S)}|\mathbf{X}) &=&\prod\limits_{s=1}^{S}g^{(s)}(Z^{(s)}|%
\mathbf{X})  \notag \\
&\propto &\sum_{\theta ^{(1)}\in \Theta ^{(1)}}...\sum_{\theta ^{(S)}\in
\Theta ^{(S)}}\prod\limits_{s=1}^{S}1_{\Theta ^{(s)}(\mathcal{L(}\mathbf{X}%
))}(\theta ^{(s)})\prod\limits_{(x,\ell )\in \mathbf{X}}\prod%
\limits_{s=1}^{S}\psi _{Z_{_{\!}}^{(s)}}^{(s,\theta ^{(s)}(\ell ))}(x,\ell )
\end{eqnarray}%
Abbreviating%
\begin{equation*}
\begin{array}{cc}
Z=(Z^{(1)},...,Z^{(S)}), & \theta =(\theta ^{(1)},...,\theta ^{(S)}), \\ 
\Theta =\Theta ^{(1)}\times ...\times \Theta ^{(S)}, & \Theta (I)=\Theta
^{(1)}(I)\times ...\times \Theta ^{(S)}(I) \\ 
1_{\Theta (I)}(\theta )=\prod\limits_{s=1}^{S}1_{\Theta ^{(s)}(I)}(\theta
^{(s)}), & \psi _{\!Z\!}^{(j^{(1)},...,j^{(S)})}(x,\ell
)=\prod\limits_{s=1}^{S}\psi _{\!Z^{(s)}\!}^{(s,j^{(s)})}(x,\ell )%
\end{array}%
\end{equation*}%
the multi-sensor likelihood function has exactly the same form as that for
the single-sensor 
\begin{equation}
g(Z|\mathbf{X})\propto \sum_{\theta \in \Theta }1_{\Theta (\mathcal{L(}%
\mathbf{X}))}(\theta )\prod\limits_{(x,\ell )\in \mathbf{X}}\psi
_{Z_{_{\!}}}^{(\theta (\ell ))}(x,\ell ).
\end{equation}%
Note that since all consitituent$\ \theta ^{(1)},...,\theta ^{(S)}$ are
positive 1-1,\ $\theta $ is said to be positive 1-1.

\subsection{Generalised Label Multi-Bernoulli (GLMB)}

A GLMB density can written in the following form 
\begin{equation}
\mathbf{\pi }(\mathbf{X})=\Delta (\mathbf{X})\sum_{\xi \in \Xi
}\sum_{I\subseteq \mathbb{L}}\omega ^{(I,\xi )}\delta _{I}[\mathcal{L(}%
\mathbf{X})]\left[ p^{(\xi )}\right] ^{\mathbf{X}}.  \label{eq:delta-glmb}
\end{equation}%
where each $\xi \in \Xi $ represents a history of (multi-sensor) association
maps $\xi =(\theta _{1:k})$, each $p^{(\xi )}(\cdot ,\ell )$ is a
probability density on $\mathbb{X}$, and each $\omega ^{(I,\xi )}$ is
non-negative with $\sum_{\xi \in \Xi }\sum_{I\subseteq \mathbb{L}}\omega
^{(I,\xi )}=1$. The cardinality distribution of a GLMB is given by%
\begin{equation}
\Pr (\left\vert \mathbf{X}\right\vert \text{=}n)=\sum_{\xi \in \Xi
}\sum_{I\subseteq \mathbb{L}}\delta _{n}\left[ \left\vert I\right\vert %
\right] \omega ^{(I,\xi )},  \label{eq:GLMBCard}
\end{equation}%
while, the existence probability and probability density of track $\ell \in 
\mathbb{L}$ are respectively\allowdisplaybreaks%
\begin{align}
r(\ell )& =\sum_{\xi \in \Xi }\sum_{I\subseteq \mathbb{L}}1_{I}(\ell )\omega
^{(I,\xi )}, \\
p(x,\ell )& =\frac{1}{r(\ell )}\sum_{\xi \in \Xi }\sum_{I\subseteq \mathbb{L}%
}1_{I}(\ell )\omega ^{(I,\xi )}p^{(\xi )}(x,\ell ).
\end{align}

Given the GLMB density (\ref{eq:delta-glmb}), an intuitive multi-object
estimator is the \emph{multi-Bernoulli estimator}, which first determines
the set of labels $L\subseteq $ $\mathbb{L}$ with existence probabilities
above a prescribed threshold, and second the MAP/mean estimates from the
densities $p(\cdot ,\ell ),\ell \in L$, for the states of the objects. A
popular estimator is a suboptimal version of the Marginal Multi-object
Estimator \cite{Mah07}, which first determines the pair $(L,\xi )$ with the
highest weight $\omega ^{(L,\xi )}$ such that $\left\vert L\right\vert $
coincides with the MAP cardinality estimate, and second the MAP/mean
estimates from $p^{(\xi )}(\cdot ,\ell ),\ell \in L$, for the states of the
objects.

\subsection{Multi-Sensor GLMB Recursion}

The GLMB filter is an analytic solution to the Bayes single-sensor
multi-object filter, under the standard multi-object dynamic and observation
models \cite{VoGLMB13}. Since the multi-sensor likelihood function has the
same form as single-sensor case, it follows from \cite{VVH16} that given the
filtering density (\ref{eq:delta-glmb}) at time $k$, the filtering density
at time $k+1$ is given by \allowdisplaybreaks%
\begin{equation}
\mathbf{\pi }_{\!+\!}(\mathbf{X})\propto \Delta (\mathbf{X}%
)\sum\limits_{I_{\!},\xi ,I_{\!+\!},\theta _{\!+\!\!}}\omega ^{(I,\xi
)}\omega _{Z_{+}}^{(_{\!}I_{\!},\xi ,I_{\!+\!},\theta _{\!+\!})}\delta
_{_{\!}I_{+}}[\mathcal{L}(_{\!}\mathbf{X}_{\!})]\!\left[ p_{Z_{+}}^{(_{\!}%
\xi ,\theta _{\!+\!})}{}\right] ^{\!\mathbf{X}}  \label{eq:GLMB_joint0}
\end{equation}%
where $I\in \mathcal{F}(\mathbb{L})$, $\xi \in \Xi $, $I_{+}\in \mathcal{F}(%
\mathbb{L}_{+})$, $\theta _{+}\in {\Theta }_{+}(I_{+})$, and%
\allowdisplaybreaks%
\begin{eqnarray}
\!\!\!\!\!\!\omega _{Z_{_{\!}+}}^{(_{\!}I_{\!},\xi ,I_{\!+\!},\theta
_{\!+\!})}\!\!\! &=&\!\!\!1_{{\Theta }_{\!+\!}(I_{+})}(\theta _{\!+\!})\!%
\left[ 1-\bar{P}_{S}^{(\xi )}\right] ^{\!I\!-I_{\!+}}\!\left[ \bar{P}%
_{S\!}^{(\xi )}\right] ^{\!I\cap I_{+\!}}\left[ 1-r_{B\!,+}\right] ^{\mathbb{%
B}_{\!+\!}-I_{\!+}}r_{B\!,+}^{\mathbb{B}_{\!{+}}\cap I_{+\!}}\!\left[ \bar{%
\psi}_{_{\!}Z_{_{\!}+}}^{(_{\!}\xi ,\theta _{_{\!}+\!})}\right] ^{I_{+}}
\label{eq:GLMB_joint1} \\
\!\!\!\!\!\!\bar{P}_{S\!}^{(\xi )}(\ell )\!\!\! &=&\!\!\!\left\langle
p^{(\xi )\!}(\cdot ,\ell ),P_{S}(\cdot ,\ell )\right\rangle
\label{eq:GLMB_joint2} \\
\!\!\!\!\!\!\bar{\psi}_{_{\!}Z_{_{\!}+}}^{(\xi ,\theta _{+\!})}(\ell
_{_{\!}+\!})\!\!\! &=&\!\!\!\left\langle \bar{p}_{+}^{(\xi )}(\cdot ,\ell
_{_{\!}+}),\psi _{_{\!}Z_{_{\!}+}\!}^{(\theta _{_{\!}+}(\ell
_{_{\!}+}))}(\cdot ,\ell _{_{\!}+})\right\rangle  \label{eq:GLMB_joint3} \\
\!\!\!\!\!\!\bar{p}_{+}^{(\xi )_{\!}}(x_{_{\!}+},\ell _{_{\!}+\!})\!\!\!
&=&\!\!\!1_{\mathbb{L}_{\!}}(\ell _{_{\!}+\!})\frac{\!\left\langle
P_{S}(\cdot ,\ell _{_{\!}+\!})f_{_{\!}+\!}(x_{_{\!}+}|\cdot ,\ell
_{_{\!}+\!}),p^{(\xi )}(\cdot ,\ell _{_{\!}+\!})\right\rangle }{\bar{P}%
_{S}^{(\xi )}(\ell _{_{\!}+})}+1_{\mathbb{B}_{+}}\!(\ell
_{_{\!}+_{\!}})p_{B,+_{_{\!}}}(x_{_{\!}+},\ell _{_{\!}+})
\label{eq:GLMB_joint4} \\
\!\!\!\!\!\!p_{Z_{_{\!}+}}^{(\xi _{\!},\theta _{\!+\!})\!}(x_{_{\!}+},\ell
_{_{\!}+\!})\!\!\! &=&\!\!\!\frac{\bar{p}_{+}^{(\xi )}(x_{_{\!}+},\ell
_{_{\!}+})\psi _{Z_{+}}^{(\theta _{_{\!}+}(\ell _{_{\!}+}))}(x_{_{\!}+},\ell
_{_{\!}+})}{\bar{\psi}_{Z_{+}}^{(\xi ,\theta _{_{\!}+})}(\ell _{_{\!}+})}.
\label{eq:GLMB_joint5}
\end{eqnarray}

Observe that (\ref{eq:GLMB_joint0}) does indeed takes on the same form as (%
\ref{eq:delta-glmb}) when rewritten as a sum over $I_{\!+},\xi ,\theta
_{\!+} $ with weights 
\begin{equation}
\omega _{+}^{(I_{\!+\!},\xi ,\theta _{\!+\!})}\propto \sum\limits_{I}\omega
^{(I,\xi )}\omega _{Z_{_{\!}+}}^{(_{\!}I_{\!},\xi ,I_{\!+\!},\theta
_{\!+\!})}.  \label{eq:GLMB_joint6}
\end{equation}%
Hence at the next iteration we only propagate forward the components $%
(I_{\!+\!},\xi ,\theta _{\!+\!})$ with weights $\omega _{+}^{(I_{\!+\!},\xi
,\theta _{\!+\!})}$.

The number of components in the $\delta $-GLMB filtering density grows
exponentially with time, and needs to be truncated at every time step,
ideally, by retaining those with largest weights since this minimizes the $%
L_{1}$ approximation error \cite{VVP_GLMB13}.

\section{Multi Sensor GLMB Implementation}

In this section we consider the truncation of the $\delta $-GLMB filtering
density (\ref{eq:GLMB_joint0}) by sampling components $(_{\!}I_{\!},\xi
,I_{\!+\!},\theta _{\!+\!})$ from some discrete probability distribution $%
\pi $. To ensure that mostly high-weight components are sampled, $\pi $
should be constructed so that only valid components have positive
probabilities, and those with high weights are more likely to be chosen than
those with low weights. A natural choice to set $\pi (I,\xi )\propto \omega
^{(I,\xi )}$ and $\pi (I_{+},\theta _{+}|I,\xi )\propto \omega
_{Z_{_{\!}+}}^{(I,\xi ,I_{+},\theta _{+})}$ so that 
\begin{equation}
\pi (I,\xi ,I_{+},\theta _{+})\propto \omega ^{(I,\xi )}\omega
_{Z_{_{\!}+}}^{(I,\xi ,I_{+},\theta _{+})}.  \label{eq:jointpi}
\end{equation}%
To draw $H_{+}^{\max }$ samples from $\pi $ , we first sample $%
\{(I^{(h)},\xi ^{(h)})\}_{h=1}^{H_{+}^{\max }}$ from $\pi (I,\xi )\propto
\omega ^{(I,\xi )}$, and then for each distinct sample $(I^{(h)},\xi ^{(h)})$
with $T_{+}^{(h)}$ copies, we draw $T_{+}^{(h)}$ samples $%
(I_{+}^{(h,t)},\theta _{+}^{(h,t)})$ from $\pi (I_{+},\theta
_{+}|I^{(h)},\xi ^{(h)})$. In the following subsections we present an
algorithm for sampling from $\pi (I_{+},\theta _{+}|I^{(h)},\xi ^{(h)})$.

\subsection{Truncation by Gibbs Sampling}

\subsubsection{The Target Distribution}

This subsection formulates the target distribution $\pi (I_{+},\theta
_{+}|I,\xi )$ for the Gibbs sampler.

Consider a fixed component $(I,\xi ,)$ of the $\delta $-GLMB filtering
density at time $k$, and a fixed measurement set $Z_{+}$ at time $k+1$.
Specifically, we enumerate $Z_{+}^{(s)}=\{z_{1:M^{(s)}}\}$, $I=\{\ell
_{1:R}\}$, and in addition $\mathbb{B}_{{+}\!}=\{\ell _{R+1:P}\}$. The goal
is to find a set of pairs $(I_{+},\theta _{+})\in \mathcal{F}(\mathbb{L}%
_{+})\times \Theta _{+}(I_{+})$ with significant $\omega
_{Z_{_{\!}+}}^{(I,\xi ,I_{+},\theta _{+})}$.

For each pair $(I_{+},\theta _{+})\in \mathcal{F}(\mathbb{L}_{+})\times
\Theta _{+}(I_{+})$, we define the array 
\begin{equation}
\gamma =\left[ 
\begin{array}{cccc}
\gamma _{1}^{(1)} & \gamma _{1}^{(2)} & \cdots & \gamma _{1}^{(S)} \\ 
\gamma _{2}^{(1)} & \gamma _{2}^{(2)} & \cdots & \gamma _{2}^{(S)} \\ 
\vdots & \vdots & \ddots & \vdots \\ 
\gamma _{P}^{(1)} & \gamma _{P}^{(2)} & \cdots & \gamma _{P}^{(S)}%
\end{array}%
\right] \in \left( \{-1\}^{S}\uplus \{0:M^{(1)}\}\times ...\times
\{0:M^{(S)}\}\right) ^{P}  \label{eq:gamma}
\end{equation}%
by 
\begin{equation*}
\gamma _{i}^{(s)}=\left\{ 
\begin{array}{ll}
\theta _{+}^{(s)}(\ell _{i}), & \text{if }\ell _{i}\in I_{+} \\ 
-1, & \text{otherwise}%
\end{array}%
\right.
\end{equation*}%
The $i$th row of $\gamma $ is denoted as $\gamma _{i}$.

Note the distinction between spaces $\left( \{-1\}^{S}\uplus
\{0:M^{(1)}\}\times ...\times \{0:M^{(S)}\}\right) ^{P}$ and $\left(
\{-1:M^{(1)}\}\times ...\times \{-1:M^{(S)}\}\right) ^{P}$: for any array $%
\gamma $ in the former, if $\gamma _{i}^{(s)}=-1$, then $\gamma _{i}$,
consist of entirely -1's. It is clear that $\gamma $ inherits, from $\theta
_{+}$, the \emph{positive 1-1} property, i.e., for each $s$ there are no
distinct $i$, $i^{\prime }\in \{1$:$P\}$ with $\gamma _{i}^{(s)}\!=\!\gamma
_{i^{\prime }}^{(s)}>0$. The set of all positive 1-1 elements of $\left(
\{-1\}^{S}\uplus \{0:M^{(1)}\}\times ...\times \{0:M^{(S)}\}\right) ^{P}$ is
denoted by ${\Gamma }$. From $\gamma \in {\Gamma }$, we can recover $I_{+}$
and $\theta _{+}:I_{+}\rightarrow \{0:M^{(1)}\}\times ...\times
\{0:M^{(S)}\}\}$, respectively, by%
\begin{equation}
I_{+}=\{\ell _{i}\in I\cup \mathbb{B}_{\!{+}\!}:\gamma _{i}\succeq 0\}\text{
and }\theta _{+}(\ell _{i})=\gamma _{i}.  \label{eq:convert}
\end{equation}%
Thus, $1_{{\Gamma }}(\gamma )=1_{_{\!}{\Theta }_{+}(I_{+})}(\theta _{+})$,
and there is a 1-1 correspondence between the spaces $\Theta _{+}(I_{+})$
and ${\Gamma }$.

Assuming that for all $i\in \{1$:$P\}$, $\bar{P}_{S}^{(\xi )\!}(\ell
_{i})\in (0,1)$ and $\bar{P}_{_{\!}D}^{(\xi )\!}(\ell _{i})\triangleq
\left\langle \bar{p}_{+}^{(\xi )\!}(\cdot ,\ell _{i}),P_{_{\!}D}(\cdot ,\ell
_{i})\right\rangle \in (0,1)$, let 
\begin{equation}
\eta _{i}(j^{(1)},...,j^{(S)})=%
\begin{cases}
1-\bar{P}_{S}^{(\xi )\!}(\ell _{i}), & \!1\leq i\leq R,\text{ }%
(j^{(1)},...,j^{(S)})\!\prec 0\!, \\ 
\bar{P}_{S\!}^{(\xi )\!}(\ell _{i})\bar{\psi}_{Z_{+_{\!}}}^{(\xi
,j^{(1)},...,j^{(S)})\!}(\ell _{i\!}), & \!1\leq i\leq R,\text{ }%
(j^{(1)},...,j^{(S)})\!\succeq 0, \\ 
1-r_{B\!,+}(\ell _{i}), & \!R\!+\!1\leq i\leq P,\text{ }%
(j^{(1)},...,j^{(S)})\!\prec 0, \\ 
r_{B\!,+}(\ell _{i})\bar{\psi}_{Z_{+_{\!}}}^{(\xi
,j^{(1)},...,j^{(S)})\!}(\ell _{i}), & \!R\!+\!1\leq i\leq P,\text{ }%
(j^{(1)},...,j^{(S)})\!\succeq 0.%
\end{cases}
\label{eq:eta}
\end{equation}%
where 
\begin{equation}
\bar{\psi}_{Z_{+_{\!}}}^{(\xi ,j^{(1)},...,j^{(S)})\!}(\ell
_{i\!})\!=\!\left\langle \bar{p}_{+}^{(\xi )}(\cdot ,\ell _{i}),\psi
_{_{\!}Z_{+}}^{(j^{(1)},...,j^{(S)})}(\cdot ,\ell _{i})\right\rangle ,
\label{eq:eta2}
\end{equation}%
and $j^{(1)},...,j^{(S)}$ are the indices of the measurements assigned to
label $\ell _{i}$, with $j^{(s)}=0$ indicating that $\ell _{i}$ is
misdetected by sensor $s$, and $j^{(1)}=...=j^{(S)}=-1$ indicating that $%
\ell _{i}$ no longer exists (note that if a row of $\gamma $ has a negative
entry then the entire row consists of negative entries an hence $\eta
_{i}(j^{(1)},...,j^{(S)})$ is only defined for $(j^{(1)},...,j^{(S)})\!%
\succeq 0$ and $(j^{(1)},...,j^{(S)})=[-1,...,-1])$. It is implicit that $%
\eta _{i}(j^{(1)},...,j^{(S)})$ depends on the given $(I,\xi ,)$ and $Z_{+}$%
, which have been omitted for compactness. The assumptions on the expected
survival and detection probabilities, $\bar{P}_{S}^{(\xi )\!}(\ell _{i})$
and $\bar{P}_{D}^{(\xi )\!}(\ell _{i})$, eliminates trivial and ideal
sensing scenarios, as well as ensuring $\eta _{i}(j^{(1)},...,j^{(S)})>0$.

Note from (\ref{eq:convert}) that since $\theta _{+}^{(s)}(\ell _{i})$ = $%
\gamma _{i}^{(s)}$, we have $\bar{\psi}_{Z_{+}}^{(\xi ,\gamma
_{i}^{(1)},...,\gamma _{i}^{(S)})}(\ell _{i})$ = $\bar{\psi}_{Z_{+}}^{(\xi
,\theta _{+}(\ell _{i}))}(\ell _{i})$ = $\bar{\psi}_{Z_{+}}^{(\xi ,\theta
_{+})}(\ell _{i})$, hence it follows from (\ref{eq:eta}) that 
\begin{eqnarray*}
\prod\limits_{n=1}^{R}\eta _{n}(\gamma _{n})\!\! &=&\!\!\left[ 1-\bar{P}%
_{S}^{(\xi )}\right] ^{I-I_{+}}\left[ \bar{P}_{S}^{(\xi )}\bar{\psi}%
_{Z_{+}}^{(\xi ,\theta _{+})}\right] ^{I\cap I_{+}}\!, \\
\prod\limits_{n=R+1}^{P}\!\eta _{n}(\gamma _{n})\!\! &=&\!\!\left[
1-r_{B\!,+}\right] ^{\mathbb{B}_{+}-I_{+}}\left[ r_{B\!,+}\bar{\psi}%
_{Z_{+}}^{(\xi ,\theta _{+})}\right] ^{\mathbb{B}_{{+}}\cap I_{+}}\!.
\end{eqnarray*}%
Moreover, using (\ref{eq:GLMB_joint1}), we have $\omega _{Z_{+}}^{(I,\xi
,I_{+},\theta _{+})}=1_{{\Gamma }}(\gamma )\prod\limits_{i=1}^{P}\eta
_{i}(\gamma _{i})$. Consequently, \emph{sampling from } $\pi (I_{+},\theta
_{+}|I,\xi )$ $\propto \omega _{Z_{_{\!}+}}^{(I,\xi ,I_{+},\theta _{+})}$ 
\emph{is equivalent to sampling from }%
\begin{equation}
\pi (\gamma )\propto 1_{{\Gamma }}(\gamma )\prod\limits_{i=1\!}^{P}\eta
_{i}(\gamma _{i})  \label{eq:thetajoint_dis}
\end{equation}

\subsubsection{Gibbs Sampling}

Formally, the Gibbs sampler is a Markov chain with transition kernel \cite%
{Geman_Gibbs84,Cassella_Gibbs92} 
\begin{equation*}
\pi (\gamma ^{\prime }|\gamma )=\dprod\limits_{i=1}^{P}\pi _{n}(\gamma
_{n}^{\prime }|\gamma _{1:n-1}^{\prime },\gamma _{n+1:P}),
\end{equation*}%
where $\pi _{n}(\gamma _{n}^{\prime }|\gamma _{1:n-1}^{\prime },\gamma
_{n+1:P})\propto \pi (\gamma _{1:n}^{\prime },\gamma _{n+1:P})$. In other
words, given $\gamma $, the rows $\gamma _{1}^{\prime },...,\gamma
_{P}^{\prime }$ of the state at the next iterate of the chain, are
distributed according to the sequence of conditionals%
\begin{align*}
\pi _{1}(\gamma _{1}^{\prime }|\gamma _{_{\!}2:P})& \propto \pi (\gamma
_{1}^{\prime },\gamma _{_{\!}2:P}) \\
& \text{ \ }\vdots \\
\pi _{n}(\gamma _{n}^{\prime }|\gamma _{1:n-1}^{\prime },\gamma _{n+1:P})&
\propto \pi (\gamma _{1:n}^{\prime },\gamma _{n+1:P}) \\
& \text{ \ }\vdots \\
\pi _{P}(\gamma _{P}^{\prime }|\gamma _{1:P-1}^{\prime })& \propto \pi
(\gamma _{1:P}^{\prime }).
\end{align*}%
Although the Gibbs sampler is computationally efficient with an acceptance
probability of 1, it requires the conditionals $\pi _{n}(\cdot |\cdot )$, $%
n\in \{1$:$P\}$, to be easily computed and sampled from. In the following we
establish closed form expressions for the conditionals.

\begin{lemma}
\label{lemma} Let $\bar{n}=\{1$:$P\}-\{n\}$, 
\begin{equation*}
\gamma _{\bar{n}\!}=\left[ 
\begin{array}{cccc}
\gamma _{1}^{(1)} & \gamma _{1}^{(2)} & \cdots & \gamma _{1}^{(S)} \\ 
\vdots & \vdots &  & \vdots \\ 
\gamma _{n-1}^{(1)} & \gamma _{n-1}^{(2)} & \cdots & \gamma _{n-1}^{(S)} \\ 
\gamma _{n+1}^{(1)} & \gamma _{n+1}^{(2)} & \cdots & \gamma _{n+1}^{(S)} \\ 
\vdots & \vdots &  & \vdots \\ 
\gamma _{P}^{(1)} & \gamma _{P}^{(2)} & \cdots & \gamma _{P}^{(S)}%
\end{array}%
\right]
\end{equation*}%
and ${\Gamma }(\bar{n})$ be the set of all positive 1-1 $\gamma _{\bar{n}}$
(i.e. $\gamma _{\bar{n}}$ such that for each $s=1,...,S$ there are no
distinct $i,j\in \bar{n}$ with $\gamma _{i}^{(s)}\!=\!\gamma
_{j}^{(s)}\!>\!0 $). Then, for any $\gamma \in \{-1$:$M\}^{P}$, $1_{{\Gamma }%
}(\gamma )$ can be factorized as:%
\begin{equation}
1_{{\Gamma }}(\gamma )=1_{{\Gamma }(\bar{n})}(\gamma _{\bar{n}%
\!})\dprod\limits_{s=1}^{S}\dprod\limits_{i\in \bar{n}}\left(
1-1_{\{1:M^{(s)}\}}(\gamma _{n}^{(s)})\delta _{\gamma _{n}^{(s)}}[\gamma
_{i}^{(s)}]\right) .  \label{eq:LemmaGibbs}
\end{equation}
\end{lemma}

\textbf{Proof: }Note that $\gamma _{i}^{(s)}\!=\!\gamma _{j}^{(s)}\!>\!0$
iff $\delta _{\gamma _{i\!}^{(s)}}[\gamma
_{j}^{(s)}]1_{\{1:M^{(s)}\}}(\gamma _{i}^{(s)})=1$. Hence, $\gamma ^{(s)}$
is positive 1-1 iff for any distinct $i$, $j$, $\delta _{\gamma
_{i\!}^{(s)}}[\gamma _{j}^{(s)}]1_{\{1:M^{(s)}\}}(\gamma _{i}^{(s)})=0$.
Also, $\gamma ^{(s)}$ is not positive 1-1 iff there exists distinct $i$, $j$
such that $\delta _{\gamma _{i\!}^{(s)}}[\gamma
_{j}^{(s)}]1_{\{1:M^{(s)}\}}(\gamma _{i}^{(s)})=1$. Similarly, $\gamma _{%
\bar{n}\!}^{(s)}$ is positive 1-1 iff for any distinct $i$, $j\in \bar{n}$, $%
\delta _{\gamma _{i\!}^{(s)}}[\gamma _{j}^{(s)}]1_{\{1:M^{(s)}\}}(\gamma
_{i}^{(s)})=0$.

We will show that (a) if $\gamma $ is positive 1-1 then the right hand side
(RHS) of (\ref{eq:LemmaGibbs}) equates to 1, and \ (b) if $\gamma $ is not
positive 1-1, then the RHS of (\ref{eq:LemmaGibbs}) equates to 0.

To establish (a), assume that $\gamma $ is positive 1-1, then $\gamma _{\bar{%
n}\!}$ is also positive 1-1, i.e., $1_{{\Gamma }(\bar{n})}(\gamma _{\bar{n}%
\!})=1$, and for any $i\neq n$, $\delta _{\gamma _{n\!}^{(s)}}[\gamma
_{i}^{(s)}]1_{\{1:M^{(s)}\}}(\gamma _{n}^{(s)})=0$ for all $s$. Hence the
RHS of (\ref{eq:LemmaGibbs}) equates to 1.

To establish (b), assume that $\gamma $ is not positive 1-1. If $\gamma _{%
\bar{n}\!}$ is also not positive 1-1, i.e., $1_{{\Gamma }(\bar{n})}(\gamma _{%
\bar{n}\!})=0$, then the RHS of (\ref{eq:LemmaGibbs}) trivially equates to
0. It remains to show that even if $\gamma _{\bar{n}\!}$ is positive 1-1,
the RHS of (\ref{eq:LemmaGibbs}) still equates to 0. Since $\gamma $ is not
positive 1-1, there exist an $s$ and distinct $i$, $j$ such that $\delta
_{\gamma _{i\!}^{(s)}}[\gamma _{j}^{(s)}]1_{\{1:M^{(s)}\}}(\gamma
_{i}^{(s)})=1$. Further, either $i$ or $j$ has to equal $n$, because the
positive 1-1 property of $\gamma _{\bar{n}\!}$ implies that if such
(distinct) $i$, $j$, are in $\bar{n}$, then $\delta _{\gamma
_{i\!}^{(s)}}[\gamma _{j}^{(s)}]1_{\{1:M^{(s)}\}}(\gamma _{i}^{(s)})=0$ and
we have a contradiction. Hence, there exist an $s$ and $i\neq n$ such that $%
\delta _{\gamma _{n\!}^{(s)}}[\gamma _{i}^{(s)}]1_{\{1:M^{(s)}\}}(\gamma
_{n}^{(s)})=1$, and thus the RHS of (\ref{eq:LemmaGibbs}) equates to 0. $%
\square $

\begin{proposition}
\label{marg_cond} For each $n\in \{1$:$P\}$, 
\begin{equation}
\pi _{n}(\gamma _{n}|\gamma _{\bar{n}})\propto \eta _{n}(\gamma
_{n})\dprod\limits_{s=1}^{S}\dprod\limits_{i\in \bar{n}}\left(
1-1_{\{1:M^{(s)}\}}(\gamma _{n}^{(s)})\delta _{\gamma _{n}^{(s)}}[\gamma
_{i}^{(s)}]\right) .  \label{eq:marg_cond1}
\end{equation}
\end{proposition}

\textbf{Proof}: We are interested in highlighting the functional dependence
of $\pi _{n}(\gamma _{n}|\gamma _{\bar{n}})$ on $\gamma _{n}$, while its
dependence on all other variables is aggregated into the normalizing
constant: 
\begin{equation*}
\pi _{n}(\gamma _{n}|\gamma _{\bar{n}})\triangleq \frac{\pi (\gamma )}{\pi
(\gamma _{\bar{n}})}\propto \pi (\gamma )\propto 1_{{\Gamma }}(\gamma
)\prod\limits_{j=1}^{P}\eta _{j}(\gamma _{j})=\eta _{n}(\gamma _{n})1_{{%
\Gamma }}(\gamma )\prod\limits_{j\in \bar{n}}\eta _{j}(\gamma _{j}).
\end{equation*}%
Factorizing $1_{{\Gamma }}(\gamma )$ using Lemma \ref{lemma}, gives 
\begin{eqnarray*}
\pi _{n}(\gamma _{n}|\gamma _{\bar{n}}) &\propto &\eta _{_{\!}n_{\!}}(\gamma
_{n})_{\!\!}\dprod\limits_{s=1}^{S}\dprod\limits_{i\in \bar{n}}\left(
1-1_{\{1:M^{(s)}\}}(\gamma _{n}^{(s)})\delta _{\gamma _{n}^{(s)}}[\gamma
_{i}^{(s)}]\right) 1_{{\Gamma }\!(\bar{n})\!}(\gamma _{\!\bar{n}%
})\!\prod\limits_{j\in \bar{n}}\eta _{j}(\gamma _{j}) \\
&\propto &\eta _{_{\!}n_{\!}}(\gamma
_{n})_{\!\!}\dprod\limits_{s=1}^{S}\dprod\limits_{i\in \bar{n}}\left(
1-1_{\{1:M^{(s)}\}}(\gamma _{n}^{(s)})\delta _{\gamma _{n}^{(s)}}[\gamma
_{i}^{(s)}]\right) .\text{ \ \ \ \ \ \ \ \ \ \ \ \ \ \ \ \ \ \ \ }\square
\end{eqnarray*}

For $(j^{(1)},...,j^{(S)})\prec 0$, $1_{\{1:M^{(s)}\}}(j^{(s)})=0$ for all $%
s $, and Proposition~\ref{marg_cond} implies $\pi _{n}(-1,...,-1|\gamma _{%
\bar{n}})$ $\propto $ $\eta _{n}(-1,...,-1)$. On the other hand, given $%
(j^{(1)},...,j^{(S)})$ $\succeq 0$, Proposition~\ref{marg_cond} implies that 
$\pi _{n}(j^{(1)},...,j^{(S)}|\gamma _{\bar{n}})$ $\propto $ $\eta
_{n}(j^{(1)},...,j^{(S)})$, unless there is an $s$ and an $i\in \bar{n}$
with $\gamma _{i}^{(s)}=j^{(s)}>0$, in which case $\pi
_{n}(j^{(1)},...,j^{(S)}|\gamma _{\bar{n}})=0$ (because $1_{\{1:M^{(s)}\}%
\!}(j^{(s)})\delta _{\!j^{(s)}}[\gamma _{i}^{(s)}]=1$). Thus, for $%
(j^{(1)},...,j^{(S)})$ $\succeq 0$ 
\begin{equation*}
\pi _{n}(j^{(1)},...,j^{(S)}|\gamma _{\bar{n}})\propto \eta
_{n}(j^{(1)},...,j^{(S)})\dprod\limits_{s=1}^{S}\left(
1-1_{\{1:M^{(s)}\}\!}(j^{(s)})1_{\{\gamma _{1:n-1}^{(s)},\gamma
_{n+1:P}^{(s)}\}}(j^{(s)})\right) .
\end{equation*}

Hence, sampling from the conditionals $\pi _{n}$ amounts to sampling from a
categorical distribution with $1+\tprod\nolimits_{s=1}^{S}(M^{(s)}+1)$
categories. This proceedure has an $\mathcal{O}(P\tprod%
\nolimits_{s=1}^{S}M^{(s)})$ complexity since sampling from a categorical
distribution is linear in the number of categories \cite{Devroye86}. The
Gibbs sampler is summarized in Algorithm 1, and has a complexity of $%
\mathcal{O}(TP^{2}\tprod\nolimits_{s=1}^{S}M^{(s)})$.

Proposition~\ref{marg_cond} also implies that for a given a positive 1-1 $%
\gamma _{\bar{n}}$, only $\gamma _{n}\in \{-1\}^{S}\uplus
\{0:M^{(1)}\}\times ...\times \{0:M^{(S)}\}$ that does not violate the
positive 1-1 property can be generated by the conditional $\pi _{n}(\cdot
|\gamma _{\bar{n}})$, with probability proportional to $\eta _{n}(\gamma
_{n})$. Thus, starting with a positive 1-1 array, all iterates of the Gibbs
sampler are also positive 1-1. If the chain is run long enough, the samples
are effectively distributed from (\ref{eq:thetajoint_dis}) as formalized in
Proposition \ref{convergence} (the proof follows directly from Proposition 4
in \cite{VVH16}).

\bigskip

\hrule

\bigskip

\textbf{Algorithm 1 Gibbs. }

\begin{itemize}
\item \textsf{{\footnotesize {input: }}}$\gamma ^{(1)},T,S,\eta =[\eta
_{i}(j^{(1)},...,j^{(S)})]$

\item \textsf{{\footnotesize {output: }}}$\gamma ^{(1)},...,\gamma ^{(T)}$
\end{itemize}

\bigskip

\hrule

\bigskip

$P:=\mathsf{size}(\eta ,1);\quad c:=[-1;...;-1];$\quad $p:=\eta ;$

\textsf{{\footnotesize {for }}}$s=1:S$

\quad $M^{(s)}:=\mathsf{size}(\eta ,1+s)-2;$

\textsf{{\footnotesize {end}}}

\textsf{{\footnotesize {for }}}$%
(j^{(1)},...,j^{(S)})=(0,...,0):(M^{(1)},...,M^{(S)})$

\quad $c:=[c;[j^{(1)},...,j^{(S)}]];$

\textsf{{\footnotesize {end}}}

\textsf{{\footnotesize {for }}}$t=2:T$

\quad $\gamma ^{\prime }:=[$ $];$

\quad \textsf{{\footnotesize {for }}}$n=1:P$

\quad \quad \textsf{{\footnotesize {for }}}$%
(j^{(1)},...,j^{(S)})=(0,...,0):(M^{(1)},...,M^{(S)})$

\quad \quad \quad $p_{n}(j^{(1)},...,j^{(S)}):=\eta
_{n}(j^{(1)},...,j^{(S)})\dprod\limits_{s=1}^{S}(1-1_{\{1:M^{(s)}\}%
\!}(j^{(s)})1_{\{\gamma _{1:n-1}^{\prime },\gamma
_{n+1:P}^{(t-1)}\}}(j^{(s)}));$

\quad \quad \textsf{{\footnotesize {end}}}

\quad \quad $\gamma _{n}^{\prime }\sim \mathsf{Categorical}(c,p_{n});$\quad $%
\gamma ^{\prime }:=[\gamma ^{\prime };\gamma _{n}^{\prime }];$

\quad \textsf{{\footnotesize {end}}}

\quad $\gamma ^{(t)}=\gamma ^{\prime };$

\textsf{{\footnotesize {end}}}

\bigskip

\hrule

\bigskip

\begin{proposition}
\label{convergence} Starting from any initial state in ${\Gamma }$, the
Gibbs sampler defined by the family of conditionals (\ref{eq:marg_cond1})
converges to the target distribution (\ref{eq:thetajoint_dis}) at an
exponential rate. More concisely, let $\pi ^{j}$ denote the $j$th power of
the transition kernel, then 
\begin{equation*}
\max_{\gamma ,\gamma ^{\prime }\in {\Gamma }}(|\pi ^{j}(\gamma ^{\prime
}|\gamma )-\pi (\gamma ^{\prime })|)\leq (1-2\beta )^{\left\lfloor \frac{j}{2%
}\right\rfloor },
\end{equation*}%
where $\beta \triangleq \min_{\gamma ,\gamma ^{\prime }\in {\Gamma }}\pi
^{2}(\gamma ^{\prime }|\gamma )>0$ is the least likely 2-step transition
probability.
\end{proposition}

The proposed Gibbs sampler has a short burn-in period due to its exponential
convergence rate. More importantly, since we are not using the samples to
approximate (\ref{eq:thetajoint_dis}) as in an MCMC inference problem, it is
not necessary to discard burn-in and wait for samples from the stationary
distribution. For the purpose of approximating the GLMB filtering density,
each distinct sample constitutes one term in the approximant, and reduces
the $L_{1}$ approximation error by an amount proportional to its weight.
Hence, regardless of their distribution, all distinct samples can be used,
the larger the weights, the smaller the $L_{1}$ error between the
approximant and the true GLMB. Note that this is also called a block Gibbs
sampler since for each row we are sampling from the joint distribution of
the elements of the row.

\subsection{Multi-Sensor Joint Prediction and Update Implementation\label%
{subsec:implementation}}

A GLMB of the form (\ref{eq:delta-glmb}) is completely characterized by
parameters $(\omega ^{(I,\xi )},p^{(\xi )})$, $(I,\xi )\in \mathcal{F}\!(%
\mathbb{L})\!\times \!\Xi $, which can be enumerated as $\{(I^{(h)},\xi
^{(h)},\omega ^{(h)},p^{(h)})\}_{h=1}^{H}$, where%
\begin{equation*}
\omega ^{(h)}\triangleq \omega ^{(I^{(h)},\xi ^{(h)})},\;p^{(h)}\triangleq
p^{(\xi ^{(h)})}.
\end{equation*}%
Since the GLMB (\ref{eq:delta-glmb}) can now be rewritten as 
\begin{equation*}
\mathbf{\pi }(\mathbf{X})=\Delta (\mathbf{X})\sum\limits_{h=1}^{H}\omega
^{(h)}\delta _{I^{(h)}}[\mathcal{L(}\mathbf{X})]\left[ p^{(h)}\right] ^{%
\mathbf{X}},
\end{equation*}%
and implementing the GLMB filter amounts to propagating forward the
parameter set $\{(I^{(h)},\omega ^{(h)},p^{(h)})\}_{h=1}^{H}.$ Note that to
be consistent with the indexing by $h$ instead of $(I,\xi )$, we abbreviate%
\begin{eqnarray}
\bar{P}_{S}^{(h)}(\ell _{_{\!}i}) &\triangleq &\bar{P}_{_{\!}S\!}^{(\xi
^{(h)})}(\ell _{_{\!}i}),\text{ \ \ }\bar{p}_{+}^{(h)_{\!}}(x,\ell
_{_{\!}i})\triangleq \bar{p}_{+}^{(\xi ^{(h)})_{\!}}(x,\ell _{_{\!}i\!}),%
\text{ \ \ }\bar{\psi}_{Z_{+}}^{(h,j^{(1)},...,j^{(S)})}(\ell
_{_{\!}i})\triangleq \bar{\psi}_{Z_{+}}^{(\xi
^{(h)},j^{(1)},...,j^{(S)})}(\ell _{_{\!}i})  \notag \\
\eta _{i}^{(h)}\left( j^{(1)},...,j^{(S)}\right) &\triangleq &%
\begin{cases}
1-\bar{P}_{S}^{(h)\!}(\ell _{i}), & \!\ell _{i\!}\in I^{(h)},\text{ }%
(j^{(1)},...,j^{(S)})\!\prec 0\!\!, \\ 
\bar{P}_{S}^{(h)}(\ell _{i})\bar{\psi}_{Z_{+}}^{(h,j^{(1)},...,j^{(S)})\!}(%
\ell _{i\!}), & \!\ell _{i\!}\in I^{(h)},\text{ }(j^{(1)},...,j^{(S)})\!%
\succeq 0, \\ 
1-r_{B\!,+}(\ell _{i}), & \!\ell _{i\!}\in \mathbb{B}_{+},\text{ }%
(j^{(1)},...,j^{(S)})\!\prec 0\!, \\ 
r_{B\!,+}(\ell _{i})\bar{\psi}_{Z_{+}}^{(h,j^{(1)},...,j^{(S)})\!}(\ell
_{i}), & \!\ell _{i\!}\in \mathbb{B}_{+},\text{ }(j^{(1)},...,j^{(S)})\!%
\succeq 0.%
\end{cases}
\label{eq:eta_h}
\end{eqnarray}

The procedure for computing the parameter set $\{(I_{+}^{(h_{+})},\omega
_{+}^{(h_{+})},p_{+}^{(h_{+})})\}_{h_{+}=1}^{H_{+}}$ at the next time
(Algorithm~2) is the same as that of the single-sensor case, with the 1-1
vectors replaced by 1-1 arrays. Note that $\{\}$ denotes a MATLAB\ cell
array of (non-unique) elements.\textbf{\ }There are three main steps in one
iteration of the GLMB filter.

First, the Gibbs sampler is used to generate the \emph{auxiliary vectors} $%
\gamma ^{(h,t)}$, $h=1$:$H$, $t=1$:$\tilde{T}_{+}^{(h)}$, with the most
significant weights $\omega _{+}^{(h,t)}$.

\bigskip

\hrule

\medskip

\textbf{Algorithm 2. Multi-Sensor Joint Prediction and Update}

\begin{itemize}
\item {\footnotesize \textsf{input: }}$\{(I^{(h)},\omega
^{(h)},p^{(h)})\}_{h=1}^{H}$, $Z_{+}$, $H_{+}^{\max }$,

\item {\footnotesize \textsf{input: }}$\{(r_{\!B\!,+}^{(\ell
)},p_{B\!,+}^{(\ell )})\}_{\ell \in \mathbb{B}_{\!+}}$, $P_{S}$, $%
f_{\!+\!}(\cdot |\cdot )$, $\{(\kappa _{+}^{(s)}$, $P_{D\!,+}^{(s)}$, $%
g_{+\!}^{(s)}(\cdot |\cdot ))\}_{s=1}^{S}$,

\item \textsf{{\footnotesize {output: }}}$\{(I_{+}^{(h_{+})},\omega
_{+}^{(h_{+})},p_{+}^{(h_{+})})\}_{h_{+}=1}^{H_{+}}$
\end{itemize}

\medskip

\hrule

\medskip

\textsf{{\footnotesize {sample counts }}}$[T_{+}^{(h)}]_{h=1}^{H}$\textsf{%
{\footnotesize {\ \ from\textsf{\ a multinomial} distribution with
parameters }}}$H_{+}^{\max }$\textsf{{\footnotesize {\ trials and weights }}}%
${[}\omega ^{(h)}]_{h=1}^{H}$

\textsf{{\footnotesize {for}}}\textsf{\ }$h=1:H$

\quad \textsf{{\footnotesize {initialize }}}$\gamma ^{(h,1)}$

\quad \textsf{{\footnotesize {compute }}}$\eta ^{(h)}$ \ \textsf{%
{\footnotesize {using }}}(\ref{eq:eta_h})

\quad $\{\gamma ^{(h,t)}\}_{_{t=1}}^{\tilde{T}_{+}^{(h)}}:=\mathsf{%
Unique(Gibbs}(\gamma ^{(h,1)},T_{+}^{(h)},S,\eta ^{(h)}));$

\quad \textsf{{\footnotesize {for }}}$t=1:\tilde{T}_{+}^{(h)}$

\quad \quad \textsf{{\footnotesize {compute }}}$I_{+}^{(h,t)}$\textsf{%
{\footnotesize {\ from }}}$I^{(h)}$\textsf{{\footnotesize {\ and }}}$\gamma
^{(h,t)}$\textsf{{\footnotesize {\ using }}}(\ref{eq:Iplus})

\quad \quad \textsf{{\footnotesize {compute }}}$\omega _{+}^{(h,t)}$\textsf{%
{\footnotesize {\ from }}}$\omega ^{(h)}$\textsf{{\footnotesize {\ and }}}$%
\gamma ^{(h,t)}$\textsf{{\footnotesize {\ using }}}(\ref{eq:wplus})

\quad \quad \textsf{{\footnotesize {compute }}}$p_{+}^{(h,t)}$\ \textsf{%
{\footnotesize {from }}}$p^{(h)}$\textsf{{\footnotesize {\ and }}}$\gamma
^{(h,t)}$\textsf{{\footnotesize {\ using}}} (\ref{eq:pplus})

\quad \textsf{{\footnotesize {end}}}

\textsf{{\footnotesize {end}}}

$(\{(I_{+}^{(h_{+})}\!,p_{+}^{(h_{+})})\}_{h_{+}=1}^{H_{+}},\sim
,[U_{h,t}]):=\mathsf{Unique}(\{(I_{+}^{(h,t)}\!,p_{+}^{(h,t)})%
\}_{(h,t)=(1,1)}^{(H,\tilde{T}_{+}^{(h)})});$

\textsf{{\footnotesize {for }}}$h_{+}=1:H_{+}$

\quad $\omega _{+}^{(h_{+})}:=\sum\limits_{h,t:U_{h,t}=h_{+}}\omega
_{+}^{(h,t)};$

\textsf{{\footnotesize {end}}}

\textsf{{\footnotesize {normalize weights }}}$\{\omega
_{+}^{(h_{+})}\}_{h_{+}=1}^{H_{+}}$

\medskip

\hrule

\vspace{5pt}

\vspace{3pt}

Second, the auxiliary vectors are used to generate an intermediate set of
parameters with the most significant weights $(I^{(h)},I_{+}^{(h,t)},\omega
_{+}^{(h,t)},p_{+}^{(h,t)})$, $h=1$:$H$, $t=1$:$\tilde{T}_{+}^{(h)}$, via 
\begin{eqnarray}
I_{+}^{(h,t)} &=&\{\ell _{i}\in I^{(h)}\cup \mathbb{B}_{\!{+}\!}:\gamma
_{i}^{(h,t)}\geq 0\},  \label{eq:Iplus} \\
\omega _{+}^{(h,t)} &\propto &\omega ^{(h)}\prod\limits_{i=1}^{|I^{(h)}\cup 
\mathbb{B}_{+}|}\eta _{i}^{(h)}(\gamma _{i}^{(h,t)}),  \label{eq:wplus} \\
p_{+}^{(h,t)\!}(\cdot ,\ell _{i}) &=&\bar{p}_{+}^{(h)}(\cdot ,\ell _{i})\psi
_{Z_{+}}^{(\gamma _{i}^{(h,t)})}(\cdot ,\ell _{i})/\bar{\psi}%
_{Z_{+}}^{(h,\gamma _{i}^{(h,t)})}(\ell _{i}).  \label{eq:pplus}
\end{eqnarray}%
Computing $p_{+}^{(h,t)}(\cdot ,\ell _{i})$ (and $\eta _{i}^{(h,t)}(j)$, $%
\bar{\psi}_{Z_{+}}^{(h,j^{(1)},...,j^{(S)})\!}(\ell _{i})$, $\bar{P}%
_{_{\!}S\!}^{(h)\!}$)\ can be done via Gaussian mixture (see subsections
IV.B of \cite{VVP_GLMB13}).

Third, the intermediate parameters are marginalized via (\ref{eq:GLMB_joint6}%
) to give the new parameter set $\{(I_{+}^{(h_{+})},\omega
_{+}^{(h_{+})},p_{+}^{(h_{+})})\}_{h_{+}=1}^{H_{+}}$. Note that $U_{h,t}$
gives the index of the GLMB component at time $k+1$ that $%
(I^{(h)},I_{+}^{(h,t)},p_{+}^{(h,t)})$ contributes to.

Since we are only interested in samples that provide a good representation
of the GLMB filtering density, increased efficiency (for the same $%
H_{+}^{\max }$) can be achieved by using annealing or tempering techniques
to modify the stationary distribution so as to induce the Gibbs sampler to
seek more diverse samples \cite{GeyerThompson95}, \cite{Neal2000} (note that
the actual weights of the GLMB components are computed using the correct
parameters). One example is to increase the temperature for diversity.
Tempering with the birth model (e.g. by feeding the Gibbs sampler with a
larger birth rate) directly induces the chain to generate more components
with births. Tempering with the survival probability induces the Gibbs
sampler to generate more components with object deaths and improves track
termination. Tempering with parameters such as detection probabilities and
clutter rate induces the Gibbs sampler to generate components that reduce
the occurrence of dropped tracks.

\section{Numerical Studies\label{sec:sim}}

\subsection{3D Linear Gaussian Scenario}

A 3D linear Gaussian scenario with 3 independent sensors is considered. An
unknown and time varying number objects appear (up to 10 simultaneously in
total) with births, deaths and crossings. Individual object kinematics are
described by a 6D state vector\ of position and velocity (in the $x$, $y$, $%
z $ directions respectively) that follows a constant velocity model with
sampling period of\ $1s$, and process noise standard deviation $\sigma _{\nu
}=5m/s^{2}$. The survival probability is $P_{S}=0.99$, and the birth model
is an LMB with parameters $\{r_{B,k}(\ell _{i}),p_{B,k}(\ell
_{i})\}_{i=1}^{4}$, where $\ell _{i}=(k,i)$, $r_{B,k}(\ell _{i})=0.03$, and $%
p_{B}(x,\ell _{i})=\mathcal{N}(x;m_{B}^{(i)},P_{B})$ with%
\begin{equation*}
\begin{array}{ll}
m_{B}^{(1)}=[0,0,0,0,0,0]^{T}, & m_{B}^{(2)}=[400,0,-600,0,200,0]^{T}, \\ 
m_{B}^{(3)}=[-800,0,-200,0,-400,0]^{T}, & 
m_{B}^{(3)}=[-200,0,800,0,600,0]^{T},%
\end{array}%
\end{equation*}

\begin{equation*}
P_{B}=\mathrm{diag}([10,10,10,10,10,10]^{T})^{2}.
\end{equation*}

For the entire scenario duration, 3 independent sensors are deployed. Each
produces 3D observations in the form of noisy position vectors on the region 
$[-1000,1000]m\times \lbrack -1000,1000]m\times \lbrack -1000,1000]m$.
Sensor 1 has good resolution on the $x$-axis only with respective noise
standard deviations $\sigma _{x}=10m,$ $\sigma _{y}=100m,\sigma _{z}=100m$
on each axis. Sensor 2 has good resolution on the $y$-axis only with noise
standard deviations $\sigma _{x}=100m,$ $\sigma _{y}=10m,\sigma _{z}=100m$
on each axis. Sensor 3 has good resolution on the $z$-axis only noise
standard deviations $\sigma _{x}=100m,$ $\sigma _{y}=100m,\sigma _{z}=10m$
on each axis. All sensors have detection probability $P_{D}=0.66$\ and
uniform Poisson false alarms with an average rate of $\lambda _{c}=20$ per
scan.

\begin{figure}[h]
\begin{center}
\resizebox{80mm}{!}{\includegraphics{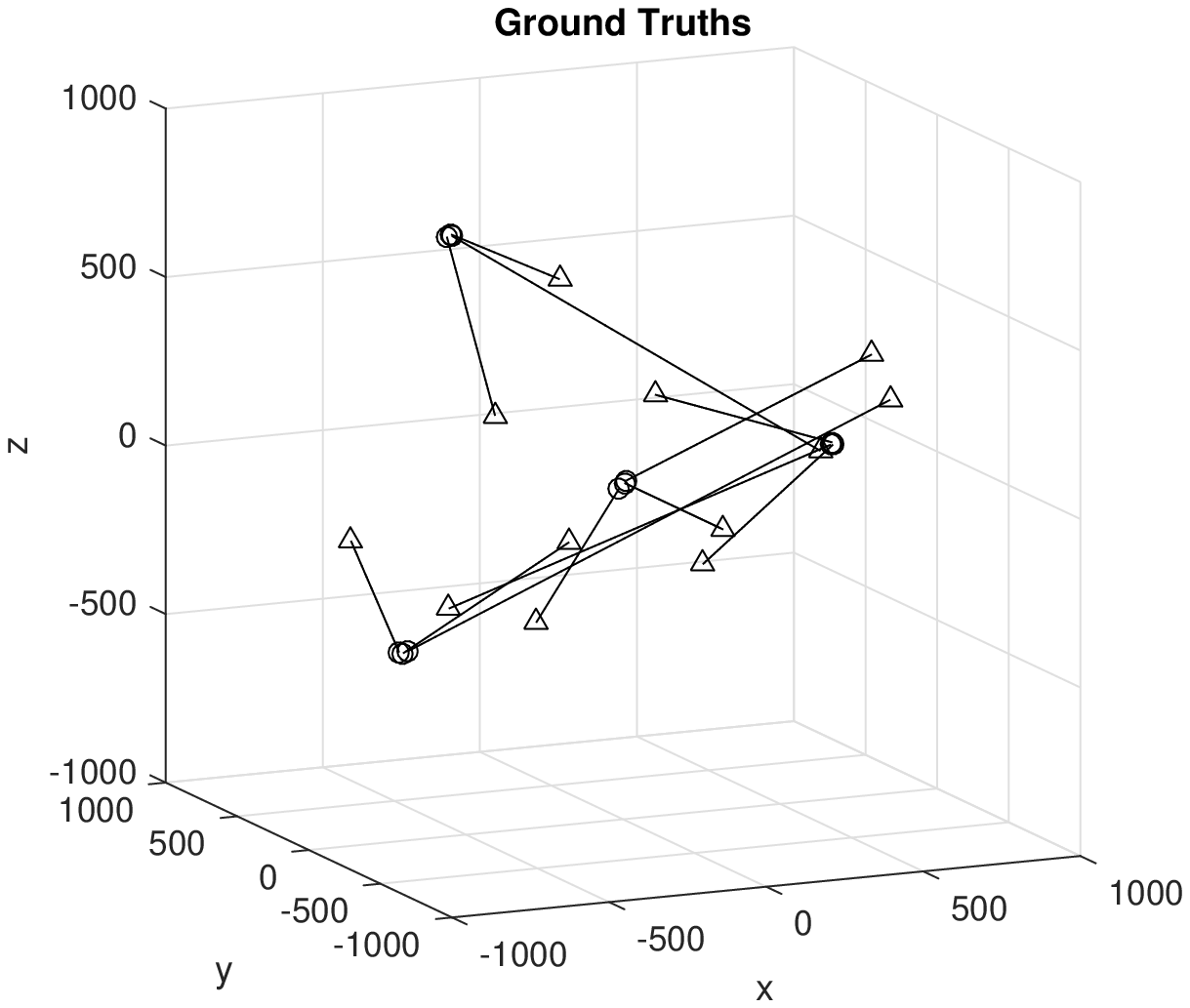}}
\end{center}
\caption{Ground truths in 3D space}
\end{figure}

\begin{figure}[h]
\begin{center}
\resizebox{80mm}{!}{\includegraphics{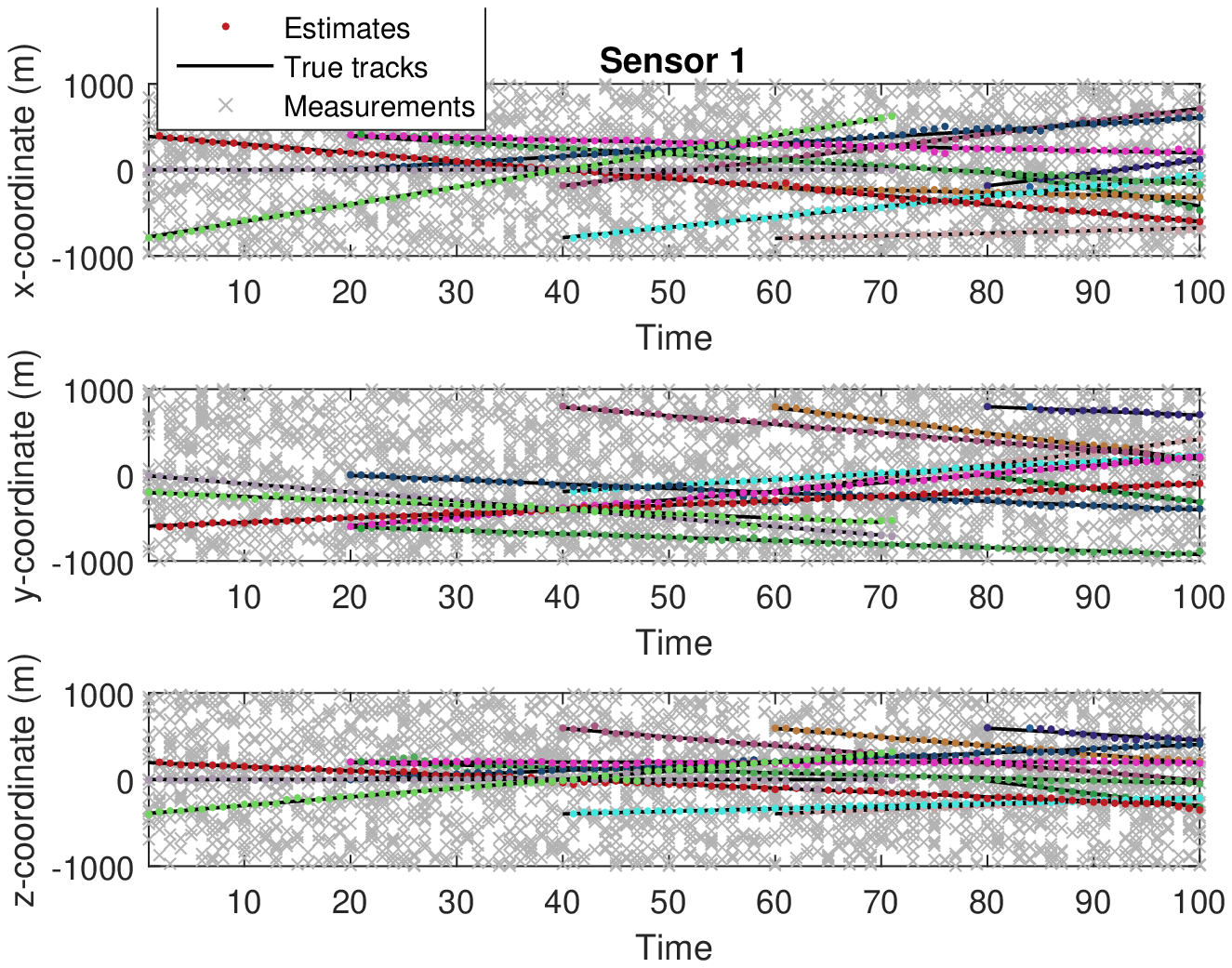}}
\end{center}
\caption{Sensor 1 measurements in $x,y,z$ coordinates versus time, also
showing true tracks and filter estimates}
\end{figure}

\begin{figure}[h]
\begin{center}
\resizebox{80mm}{!}{\includegraphics{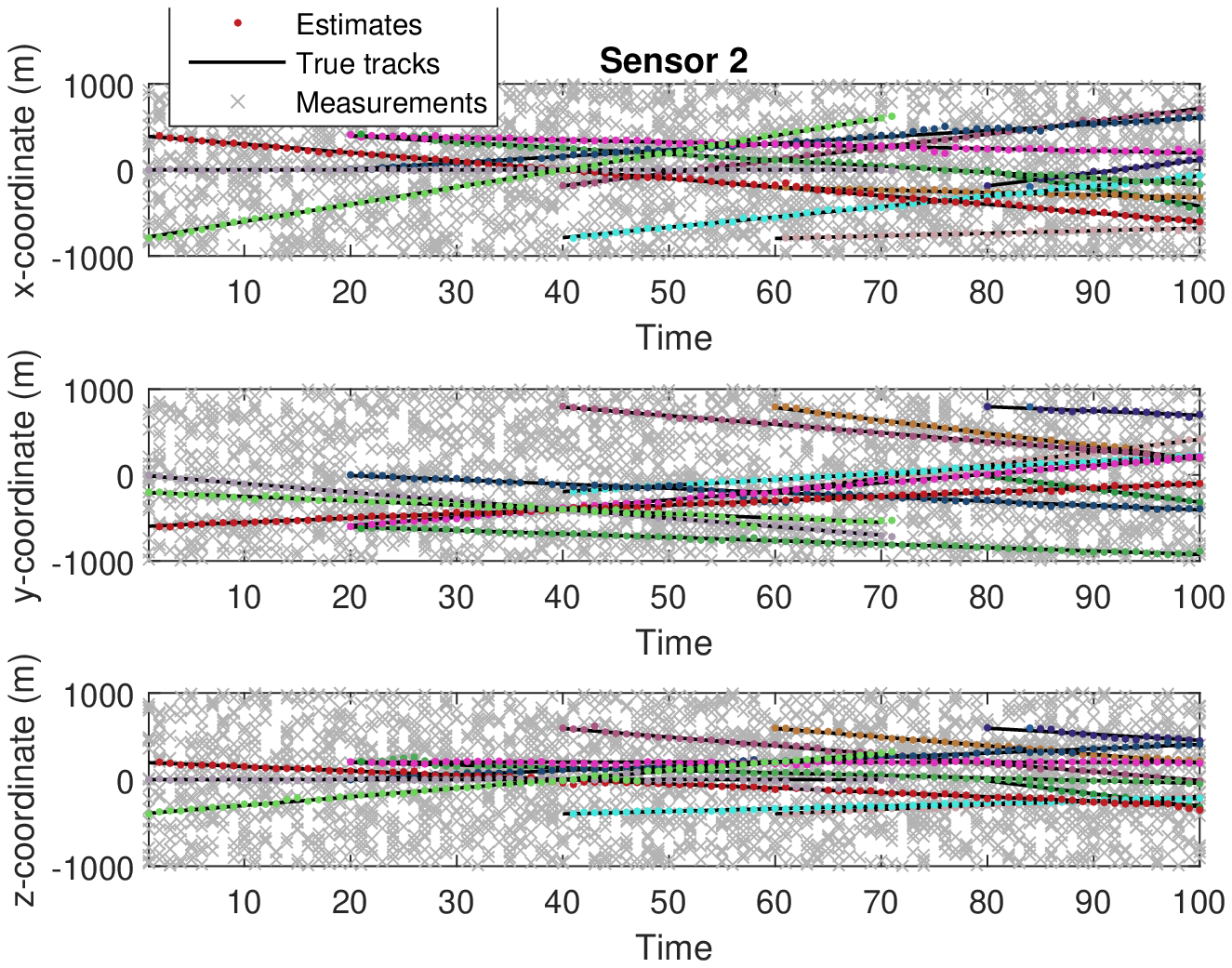}}
\end{center}
\caption{Sensor 2 measurements in $x,y,z$ coordinates versus time, also
showing true tracks and filter estimates}
\end{figure}

\begin{figure}[h]
\begin{center}
\resizebox{80mm}{!}{\includegraphics{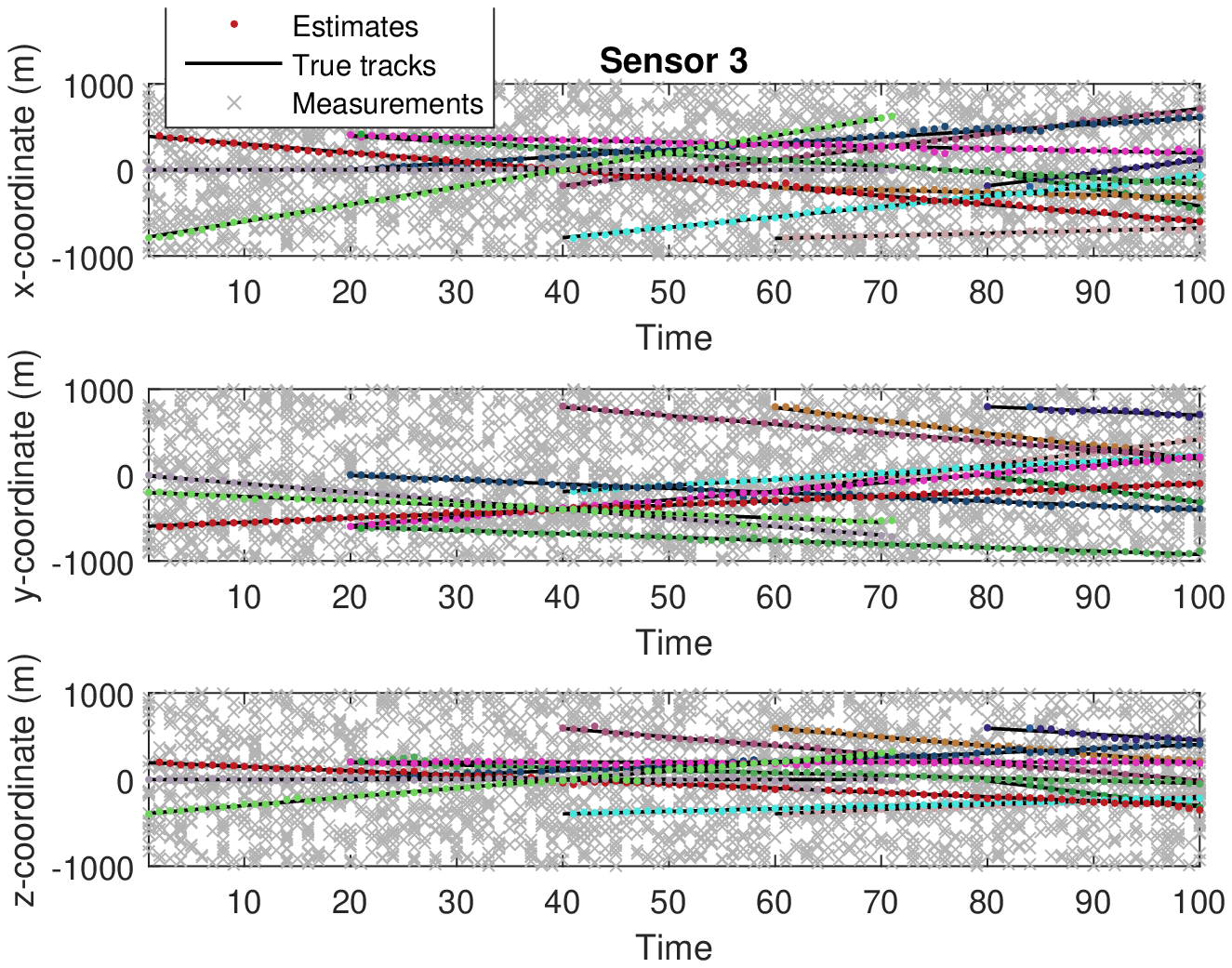}}
\end{center}
\caption{Sensor 3 measurements in $x,y,z$ coordinates versus time, also
showing true tracks and filter estimates}
\end{figure}

\begin{figure}[h]
\begin{center}
\resizebox{80mm}{!}{\includegraphics{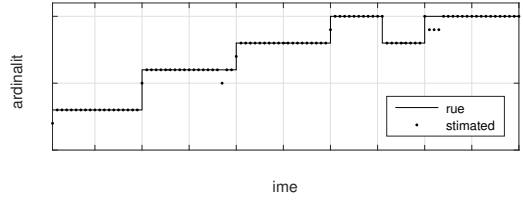}}
\end{center}
\caption{True and estimated cardinality versus time}
\end{figure}

\begin{figure}[h]
\begin{center}
\resizebox{80mm}{!}{\includegraphics{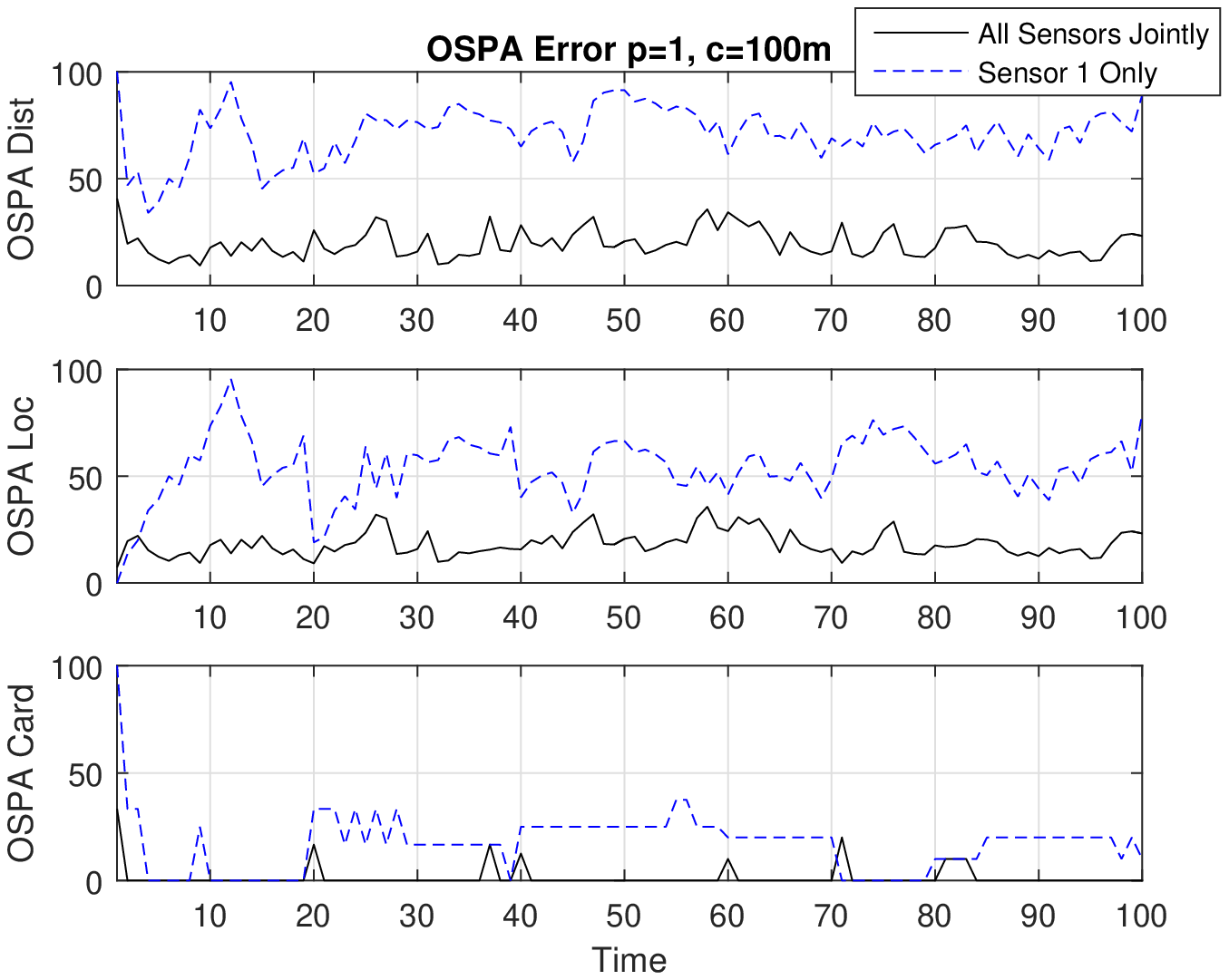}}
\end{center}
\caption{OSPA error with localization and cardinality components (p=1,
c=100m)}
\label{fig:lgospa}
\end{figure}

The multi-sensor GLMB filter is implemented via the proposed Gibbs sampling
technique. The filter is run with 10000 components and the Gibbs sampler is
tempered by taking the $3$rd root of the cost tensor corresponding to a
total of 3 sensors. Figure 1 shows the ground truths in 3D space. Figures 2,
3, 4 respectively show the measurements for sensors 1, 2, 3, in $x,y,z$
coordinates versus time, along with the ground truths and filter estimates
also superimposed on the same figures. Note that the truths and estimates
are the same for all sensors and are repeated on all sensor figures for
convenience. Figure 5 shows the true and estimated cardinality versus time
and Figure 6 shows the OSPA error ($p=1$ and $c=100m$) for the single and
multiple sensor GLMB filters. It can be seen that for the multi-sensor GLMB
filter all tracks are initiated and terminated correctly and the positions
estimates are mostly accurate. This assessment is confirmed by the OSPA
error which has a localization component consistent with the measurement
noise and peaks in the cardinality component corresponding to times of
target births and deaths.

\section{Efficient Implementation}

\subsection{Memory vs Computation}

The input to the Gibbs sampler (Algorithm 1) consists of \ $%
P(1+\tprod\nolimits_{s=1}^{S}(1+M^{(s)}))$ entries for $\eta
_{n}(j^{(1)},...,j^{(S)})$, $n=1,...,P$, and $(j^{(1)},...,j^{(S)})\in
\{-1\}^{S}\uplus \{0:M^{(1)}\}\times ...\times \{0:M^{(S)}\}$ (see (\ref%
{eq:eta})). For each $n$, the entries $\eta _{n}(j^{(1)},...,j^{(S)})$, $%
(j^{(1)},...,j^{(S)})\in \{-1\}^{S}\uplus \{0:M^{(1)}\}\times ...\times
\{0:M^{(S)}\}$ are used to construct a categorical distribution $\pi
_{n}(j^{(1)},...,j^{(S)})$ with $1+\tprod\nolimits_{s=1}^{S}(1+M^{(s)})$
categories by%
\begin{equation}
\pi _{n}(j^{(1)},...,j^{(S)})=K_{n}^{-1}\tilde{\eta}_{n}(j^{(1)},...,j^{(S)})
\end{equation}%
where%
\begin{eqnarray*}
\tilde{\eta}_{n}(j^{(1)},...,j^{(S)}) &=&\eta
_{n}(j^{(1)},...,j^{(S)})\dprod\limits_{s=1}^{S}\dprod\limits_{i\in \bar{n}%
}\left( 1-1_{\{1:M^{(s)}\}}(j^{(s)})\delta _{j^{(s)}}[\gamma
_{i}^{(s)}]\right) \\
K_{n} &=&\sum_{j^{(1)},...,j^{(S)}}\tilde{\eta}_{n}(j^{(1)},...,j^{(S)})
\end{eqnarray*}%
(see Proposition \ref{marg_cond}). The memory required for the categorical
distribution is thus of order $\mathcal{O}(\tprod\nolimits_{s=1}^{S}M^{(s)})$%
.

If we factorize%
\begin{eqnarray}
\pi _{n}(j^{(1)},...,j^{(S)}) &=&\pi
_{n}^{(S)}(j^{(S)}|j^{(S-1)},...,j^{(1)})...\pi
_{n}^{(2)}(j^{(2)}|j^{(1)})\pi _{n}^{(1)}(j^{(1)})  \notag \\
&=&\dprod\limits_{s=1}^{S}\pi _{n}^{(s)}(j^{(s)}|j^{(1)},...,j^{(s-1)}) 
\notag \\
&=&K_{n}^{-1}\dprod\limits_{s=1}^{S}\tilde{\eta}%
_{n}^{(s)}(j^{(s)}|j^{(1)},...,j^{(s-1)})
\end{eqnarray}%
then sampling $(j^{(1)},...,j^{(S)})$ from $\pi _{n}$ is equivalent to $%
j^{(1)}\sim \pi _{n}^{(1)}$, $j^{(2)}\sim $ $\pi _{n}^{(2)}(\cdot |j^{(1)})$%
, ...., $j^{(S)}\sim \pi _{n}^{(S)}(\cdot |j^{(S-1)},...,j^{(1)})$. Hence
instead of storing $1+\tprod\nolimits_{s=1}^{S}(1+M^{(s)})$ categories, we
only require $2+\max_{s}M^{(s)}$ categories.

To determine the conditionals $\tilde{\eta}%
_{n}^{(s)}(j^{(s)}|j^{(1)},...,j^{(s-1)})$ we first specify some
abbreviations. A Gaussian with mean $m$ and covariance $P$ is denoted by $%
\mathcal{N}(\cdot ;m,P)$. Given $(m$,$P)$, the Kalman updated
mean-covariance pair with measurement $j^{(s)}$ from sensor $s$, is denoted
by $(m^{(s)}(j^{(s)}),P^{(s)}(j^{(s)}))$, i.e.%
\begin{eqnarray*}
\mathcal{N}(y;m,P)\mathcal{N}(z_{j^{(s)}}^{(s)};H^{(s)}y,R^{(s)}) &=&%
\mathcal{N}(y;m^{(s)}(j^{(s)}),P^{(s)}(j^{(s)}))q^{(s)}(j^{(s)}) \\
q^{(s)}(j^{(s)}) &\triangleq &\mathcal{N}%
(z_{j^{(s)}}^{(s)};H^{(s)}m,H^{(s)}PH^{(s)\text{T}}+R^{(s)})
\end{eqnarray*}%
Suppose $(m^{(s)}(j^{(s)}),P^{(s)}(j^{(s)}))$ is further updated with
measurement $j^{(s)}$ from sensor $s$, then we denote the updated
mean-covariance pair by $%
(m^{(s,t)}(j^{(s)},j^{(t)}),P^{(s,t)}(j^{(s)},j^{(t)}))$, i.e.%
\begin{eqnarray*}
\mathcal{N}\!(y;m^{(s)}(j^{(s)}),P^{(s)}(j^{(s)}))\mathcal{N}%
(z_{j^{(t)}}^{(t)};H^{(t)}y,R^{(t)})\!\! &=&\!\!\mathcal{N}%
\!(y;m^{(s,t)}(j^{(s)}\!,j^{(t)}),P^{(s,t)}(j^{(s)}%
\!,j^{(t)}))q^{(s,t)}(j^{(s)}\!,j^{(t)}) \\
q^{(s,t)}(j^{(s)},j^{(t)})\!\! &=&\!\!\mathcal{N}%
\!(z_{j^{(t)}}^{(t)};H^{(t)}m^{(s)}(j^{(s)}),H^{(t)}P^{(s)}(j^{(s)})H^{(t)%
\text{T}}+R^{(t)})
\end{eqnarray*}%
Further, given any $\mathcal{N}\left( y;m^{(\zeta ,\ell _{n})},P^{(\zeta
,\ell _{n})}\right) $ and $\psi _{\!Z_{+}^{(s)}\!}^{(s,j^{(s)})}(y,\ell _{n})
$ we define%
\begin{eqnarray}
(\bar{m}^{(\zeta ,\ell _{n},s)}(j^{(s)}),\bar{P}^{(\zeta ,\ell
_{n},s)}(j^{(s)})) &\triangleq &\left\{ 
\begin{array}{cc}
(m^{(\zeta ,\ell _{n})},P^{(\zeta ,\ell _{n})}), & j^{(s)}=0 \\ 
(m^{(\zeta ,\ell _{n},s)}(j^{(s)}),P^{(\zeta ,\ell _{n},s)}), & j^{(s)}>0%
\end{array}%
\right.  \\
\bar{q}^{(\zeta ,\ell _{n},s)}(j^{(s)}) &\triangleq &\left\{ 
\begin{array}{cc}
1-P_{\!D}^{(s)}(\ell _{n}), & j^{(s)}=0 \\ 
q^{(\zeta ,\ell _{n},s)}(j^{(s)})P_{\!D}^{(s)}(\ell _{n})/\kappa
^{(s)}(z_{j^{(s)}}^{(s)}), & j^{(s)}>0%
\end{array}%
\right. 
\end{eqnarray}%
so that%
\begin{equation}
\mathcal{N}(y;m^{(\zeta ,\ell _{n})},P^{(\zeta ,\ell _{n})})\psi
_{\!Z_{+}^{(s)}\!}^{(s,j^{(s)})}(y,\ell _{n})=\mathcal{N}(y;\bar{m}^{(\zeta
,\ell _{n},s)}(j^{(s)}),\bar{P}^{(\zeta ,\ell _{n},s)}(j^{(s)}))\bar{q}%
^{(\zeta ,\ell _{n},s)}(j^{(s)})
\end{equation}%
Similarly we define $\bar{m}^{(\zeta ,\ell
_{n},1,...,t)}(j^{(1)}\!,...,j^{(t)})$, $\bar{P}^{(\zeta ,\ell
_{n},1,...,t)}(j^{(1)}\!,...,j^{(t)})$, and $\bar{q}^{(\zeta ,\ell
_{n},1,...,t)}(j^{(1)}\!,...,j^{(t)})$ so that%
\begin{eqnarray}
&&\!\!\!\!\!\!\!\!\!\!\!\!\!\!\!\!\!\!\!\!\!\!\!\!\!\!\!\!\!\mathcal{N}(y;%
\bar{m}^{(\zeta ,\ell _{n},1,...,t-1)}(j^{(1)},...,j^{(t-1)}),\bar{P}%
^{(\zeta ,\ell _{n},1,...,t-1)}(j^{(1)},...,j^{(t-1)}))\psi
_{\!Z_{+}^{(t)}\!}^{(t,j^{(t)})}(y,\ell _{n}) \\
&=&\mathcal{N}(y;\bar{m}^{(\zeta ,\ell _{n},1,...,t)}(j^{(1)},...,j^{(t)}),%
\bar{P}^{(\zeta ,\ell _{n},1,...,t)}(j^{(1)},...,j^{(t)}))\bar{q}^{(\xi
,\ell _{n},1,...,t)}(j^{(1)},...,j^{(t)})
\end{eqnarray}

Note that the term $\bar{\psi}_{Z_{+_{\!}}}^{(\xi
,j^{(1)},...,j^{(t)})\!}(\ell _{i\!})$ in $\eta _{n}(j^{(1)},...,j^{(t)})$
can be written as 
\begin{equation}
\bar{\psi}_{Z_{+_{\!}}}^{(\xi ,j^{(1)},...,j^{(t)})\!}(\ell _{n\!})=\int 
\bar{p}_{+}^{(\xi )}(y,\ell _{n})\prod\limits_{s=1}^{t}\psi
_{\!Z_{+}^{(s)}\!}^{(s,j^{(s)})}(y,\ell _{n})dy.  \label{eq:jointassociation}
\end{equation}%
by substituting the definition of $\psi
_{\!Z_{+}\!}^{(j^{(1)},...,j^{(t)})}(\cdot ,\ell _{n})$ into (\ref{eq:eta2}%
). Since $\bar{p}_{+}^{(\xi )}(y,\ell _{n})$ is a Gaussian, say $\bar{p}%
_{+}^{(\xi )}(y,\ell _{n})=\mathcal{N}\left( y;m^{(\xi ,\ell _{n})},P^{(\xi
,\ell _{n})}\right) $, we have%
\begin{eqnarray*}
\bar{\psi}_{Z_{+_{\!}}}^{(\xi ,j^{(1)},...,j^{(t)})\!}(\ell _{n\!}) &=&\int 
\mathcal{N}\left( y;m^{(\xi ,\ell _{n})},P^{(\xi ,\ell _{n})}\right)
\prod\limits_{s=1}^{t}\psi _{\!Z_{+}^{(s)}\!}^{(s,j^{(s)})}(y,\ell _{n})dy \\
&=&\int \mathcal{N}\left( y;m^{(\xi ,\ell _{n})},P^{(\xi ,\ell _{n})}\right)
\psi _{\!Z_{+}^{(1)}\!}^{(1,j^{(1)})}(y,\ell _{n})\prod\limits_{s=2}^{t}\psi
_{\!Z_{+}^{(s)}\!}^{(s,j^{(s)})}(y,\ell _{n})dy \\
&=&\bar{q}^{(\xi ,\ell _{n},1)}\left( j^{(1)}\right) \int \mathcal{N}\left(
y;\bar{m}^{(\xi ,\ell _{n},1)}(j^{(1)}),\bar{P}^{(\xi ,\ell
_{n},1)}(j^{(1)})\right) \prod\limits_{s=2}^{t}\psi
_{\!Z_{+}^{(s)}\!}^{(s,j^{(s)})}(y,\ell _{n})dy
\end{eqnarray*}%
Consequently, iterating we have%
\begin{eqnarray}
\bar{\psi}_{Z_{+_{\!}}}^{(\xi ,j^{(1)},...,j^{(t)})\!}(\ell _{n\!}) &=&\bar{q%
}^{(\xi ,\ell _{n},1)}(j^{(1)})\int \mathcal{N}(y;\bar{m}^{(\xi ,\ell
_{n},1)}(j^{(1)}),\bar{P}^{(\xi ,\ell _{n},1)})\prod\limits_{s=2}^{t}\psi
_{\!Z_{+}^{(s)}\!}^{(s,j^{(s)})}(y,\ell _{n})dy  \notag \\
&=&\bar{q}^{(\xi ,\ell _{n},1)}(j^{(1)})\bar{q}^{(\xi ,\ell
_{n},1,2)}(j^{(1)},j^{(2)})\times   \notag \\
&&\int \mathcal{N}(y;\bar{m}^{(\xi ,\ell _{n},1,2)}(j^{(1)},j^{(2)}),\bar{P}%
^{(\xi ,\ell _{n},1,2)}(j^{(1)},j^{(2)}))\prod\limits_{s=3}^{t}\psi
_{\!Z_{+}^{(s)}\!}^{(s,j^{(s)})}(y,\ell _{n})dy  \notag \\
&&\vdots   \notag \\
&=&\bar{q}^{(\xi ,\ell _{n},1)}(j^{(1)})\bar{q}^{(\xi ,\ell
_{n},1,2)}(j^{(1)},j^{(2)})...\bar{q}^{(\xi ,\ell
_{n},1,...,t)}(j^{(1)},...,j^{(t)})\times   \notag \\
&&\int \mathcal{N}(y;\bar{m}^{(\xi ,\ell
_{n},1,2,...t)}(j^{(1)},...,j^{(t)}),\bar{P}^{(\xi ,\ell
_{n},1,...,t)}(j^{(1)},...,j^{(t)}))dy  \notag \\
&=&\bar{q}^{(\xi ,\ell _{n},1)}(j^{(1)})\bar{q}^{(\xi ,\ell
_{n},1,2)}(j^{(1)},j^{(2)})...\bar{q}^{(\xi ,\ell
_{n},1,...,t)}(j^{(1)},...,j^{(t)})
\end{eqnarray}

Equation (\ref{eq:eta}) (reproduced here for convenience)%
\begin{equation*}
\eta _{n}(j^{(1)},...,j^{(t)})=%
\begin{cases}
1-\bar{P}_{S}^{(\xi )\!}(\ell _{n}), & \!1\leq n\leq R,\text{ }%
(j^{(1)},...,j^{(t)})\!\prec 0\!, \\ 
\bar{P}_{S\!}^{(\xi )\!}(\ell _{n})\bar{\psi}_{Z_{+_{\!}}}^{(\xi
,j^{(1)},...,j^{(t)})\!}(\ell _{n\!}), & \!1\leq n\leq R,\text{ }%
(j^{(1)},...,j^{(t)})\!\succeq 0, \\ 
1-r_{B\!,+}(\ell _{n}), & \!R\!+\!1\leq n\leq P,\text{ }%
(j^{(1)},...,j^{(t)})\!\prec 0, \\ 
r_{B\!,+}(\ell _{n})\bar{\psi}_{Z_{+_{\!}}}^{(\xi
,j^{(1)},...,j^{(t)})\!}(\ell _{n}), & \!R\!+\!1\leq n\leq P,\text{ }%
(j^{(1)},...,j^{(t)})\!\succeq 0.%
\end{cases}%
\end{equation*}%
can be written as a product of the factors 
\begin{equation}
\eta _{n}(j^{(1)})=%
\begin{cases}
1-\bar{P}_{S}^{(\xi )\!}(\ell _{n}), & \!1\leq n\leq R,\text{ }j^{(1)}=-1,
\\ 
\bar{P}_{S\!}^{(\xi )\!}(\ell _{n})\bar{q}^{(\xi ,\ell _{n},1)}(j^{(1)}), & 
\!1\leq n\leq R,\text{ }j^{(1)}\geq 0, \\ 
1-r_{B\!,+}(\ell _{n}), & \!R\!+\!1\leq n\leq P,\text{ }j^{(1)}=-1, \\ 
r_{B\!,+}(\ell _{n})\bar{q}^{(\xi ,\ell _{n},1)}(j^{(1)}), & \!R\!+\!1\leq
n\leq P,\text{ }j^{(1)}\geq 0.%
\end{cases}%
\end{equation}%
and for $t=2,...,S$%
\begin{equation*}
\eta _{n}(j^{(t)}|j^{(1)},...,j^{(t-1)})=\left\{ 
\begin{array}{cc}
1, & j^{(t)}=j^{(t-1)}=...=j^{(1)}=-1, \\ 
\bar{q}^{(\xi ,\ell _{n},1,2,...,t)}(j^{(1)},...,j^{(t)}), & 
j^{(1)},...,j^{(t)}\geq 0, \\ 
0, & \text{otherwise}%
\end{array}%
\right.
\end{equation*}%
It can be seen from the above equations that the factors $\tilde{\eta}%
_{n}^{(1)}(\cdot ),$ $\tilde{\eta}_{n}^{(2)}(\cdot |j^{(1)})$,...., $\tilde{%
\eta}_{n}^{(S)}(\cdot |j^{(S-1)},...,j^{(1)})$ can be computed on-the-fly.
However, the corresponding probability distributions $\pi _{n}^{(1)}(\cdot )$%
, $\pi _{n}^{(2)}(\cdot |j^{(1)})$,..., $\pi _{n}^{(S)}(\cdot
|j^{(S-1)},...,j^{(1)})$ involve the normalizing constants:%
\begin{eqnarray}
K_{n}^{(1)}(j^{(1)}) &\triangleq
&\sum_{j^{(2)},...,j^{(S)}}\dprod\limits_{s=2}^{S}\tilde{\eta}%
_{n}^{(s)}(j^{(s)}|j^{(1)},...,j^{(s-1)}) \\
&&\vdots  \notag \\
K_{n}^{(1,...,t)}(j^{(1)},...,j^{(t)}) &\triangleq
&\sum_{j^{(t+1)},...,j^{(S)}}\dprod\limits_{s=t+1}^{S}\tilde{\eta}%
_{n}^{(s)}(j^{(s)}|j^{(1)},...,j^{(s-1)}) \\
&&\vdots  \notag \\
K_{n}^{(1,...,S-1)}(j^{(1)},...,j^{(S-1)}) &\triangleq &\sum_{j^{(S)}}\tilde{%
\eta}_{n}^{(S)}(j^{(S)}|j^{(1)},...,j^{(S-1)})
\end{eqnarray}%
since 
\begin{eqnarray}
\pi _{n}^{(1)}(j^{(1)}) &=&\sum_{j^{(2)},...,j^{(S)}}\pi
_{n}(j^{(1)},...,j^{(S)})=\frac{1}{K_{n}}\sum_{j^{(2)},...,j^{(S)}}\dprod%
\limits_{s=1}^{S}\tilde{\eta}_{n}^{(s)}(j^{(s)}|j^{(1)},...,j^{(s-1)}) 
\notag \\
&=&\frac{1}{K_{n}}\tilde{\eta}_{n}^{(1)}(j^{(1)})\sum_{j^{(2)},...,j^{(S)}}%
\dprod\limits_{s=2}^{S}\tilde{\eta}_{n}^{(s)}(j^{(s)}|j^{(1)},...,j^{(s-1)})
\notag \\
&=&\tilde{\eta}_{n}^{(1)}(j^{(1)})\frac{K_{n}^{(1)}(j^{(1)})}{K_{n}}
\end{eqnarray}%
and 
\begin{eqnarray}
\pi _{n}^{(t)}(j^{(t)}|j^{(1)},...,j^{(t-1)}) &=&\frac{\pi
_{n}(j^{(1)},...,j^{(t-1)},j^{(t)})}{\pi _{n}(j^{(1)},...,j^{(t-1)})}=\frac{%
\sum\limits_{j^{(t+1)},...,j^{(S)}}\pi _{n}(j^{(1)},...,j^{(S)})}{%
\sum\limits_{j^{(t)},...,j^{(S)}}\pi _{n}(j^{(1)},...,j^{(S)})}  \notag \\
&=&\frac{\prod\limits_{s=1}^{t}\tilde{\eta}%
_{n}^{(s)}(j^{(s)}|j^{(1)},...,j^{(s-1)})\sum\limits_{j^{(t+1)},...,j^{(S)}}%
\prod\limits_{s=t+1}^{S}\tilde{\eta}_{n}^{(s)}(j^{(s)}|j^{(1)},...,j^{(s-1)})%
}{\prod\limits_{s=1}^{t-1}\tilde{\eta}%
_{n}^{(s)}(j^{(s)}|j^{(1)},...,j^{(s-1)})\sum\limits_{j^{(t)},...,j^{(S)}}%
\prod\limits_{s=t}^{S}\tilde{\eta}_{n}^{(s)}(j^{(s)}|j^{(1)},...,j^{(s-1)})}
\notag \\
&=&\tilde{\eta}_{n}^{(t)}(j^{(t)}|j^{(1)},...,j^{(t-1)})\frac{%
K_{n}^{(1,...,t)}(j^{(1)},...,j^{(t)})}{%
K_{n}^{(1,...,t-1)}(j^{(1)},...,j^{(t-1)})} \\
&&\vdots  \notag \\
\pi _{n}^{(S)}(j^{(S)}|j^{(1)},...,j^{(S-1)}) &=&\tilde{\eta}%
_{n}^{(S)}(j^{(S)}|j^{(1)},...,j^{(S-1)})\frac{1}{%
K_{n}^{(1,...,S-1)}(j^{(1)},...,j^{(S-1)})}
\end{eqnarray}%
The complexity of computing the normalizing constants are still of order $%
\mathcal{O}(\tprod\nolimits_{s=1}^{S}M^{(s)})$. Note that computing these
recursively as follows%
\begin{eqnarray}
K_{n}^{(1,...,S-2)}(j^{(1)},...,j^{(S-2)}) &=&\sum_{j^{(S-1)}}\tilde{\eta}%
_{n}^{(S-1)}(j^{(S-1)}|j^{(1)},...,j^{(S-2)})\sum_{j^{(S)}}\tilde{\eta}%
_{n}^{(S)}(j^{(S)}|j^{(1)},...,j^{(S-1)})  \notag \\
&=&\sum_{j^{(S-1)}}\tilde{\eta}%
_{n}^{(S-1)}(j^{(S-1)}|j^{(1)},...,j^{(S-2)})K_{n}^{(1,...,S-1)}(j^{(1)},...,j^{(S-1)})
\\
&&\vdots  \notag \\
K_{n}^{(1,...,t)}(j^{(1)},...,j^{(t)}) &=&\sum_{j^{(t+1)}}\tilde{\eta}%
_{n}^{(t+1)}(j^{(t+1)}|j^{(1)},...,j^{(t)})\sum_{j^{(t)},...,j^{(S)}}\dprod%
\limits_{s=t}^{S}\tilde{\eta}_{n}^{(s)}(j^{(s)}|j^{(1)},...,j^{(s-1)}) 
\notag \\
&=&\sum_{j^{(t+1)}}\tilde{\eta}%
_{n}^{(t+1)}(j^{(t+1)}|j^{(1)},...,j^{(t)})K_{n}^{(1,...,t+1)}(j^{(1)},...,j^{(t+1)})
\\
&&\vdots  \notag \\
K_{n}^{(1)}(j^{(1)}) &=&\sum_{j^{(2)}}\tilde{\eta}%
_{n}^{(2)}(j^{(2)}|j^{(1)})\sum_{j^{(3)},...,j^{(S)}}\dprod\limits_{s=3}^{S}%
\tilde{\eta}_{n}^{(s)}(j^{(s)}|j^{(1)},...,j^{(s-1)})  \notag \\
&=&\sum_{j^{(2)}}\tilde{\eta}%
_{n}^{(2)}(j^{(2)}|j^{(1)})K_{n}^{(1,2)}(j^{(1)},j^{(2)})
\end{eqnarray}%
saves repeated computations but still incurs an $\mathcal{O}%
(\tprod\nolimits_{s=1}^{S}M^{(s)})$ complexity.

It is possible to replace sampling from the categorical distribution $\tilde{%
\eta}_{n}(\cdot |\gamma _{1:n-1}^{\prime },\gamma _{n+1:P})$ in the (block)
Gibbs sampler by a single iteration of the Metropolis-Hastings algorithm
with target distribution $\tilde{\eta}_{n}(\cdot |\gamma _{1:n-1}^{\prime
},\gamma _{n+1:P})$. Other alternatives include adaptive rejection sampling
(ARS) \cite{GilksWild92}, adaptive rejection Metropolis sampling \cite%
{GilksBestTan95}, and others such as \cite{Ritteretal92}. While such
approach avoids the $\mathcal{O}(\tprod\nolimits_{s=1}^{S}M^{(s)})$
complexity, the Gibbs sampler may take longer to converge.

\subsection{Alternative Target Distribution}

Note that (\ref{eq:jointpi}) is not the only choice of distribution that
ensures valid components with high weights are chosen more often than those
with low weights. Instead of sampling from $\pi (I_{+},\theta _{+}|I,\xi
)\propto \omega _{Z_{_{\!}+}}^{(I,\xi ,I_{+},\theta _{+})}$, this subsection
introduces an alternative target distribution for the Gibbs sampler, which
can drastically reduce the complexity. Unique samples from the alternative
target distribution are then reweighted according to (\ref{eq:wplus}).

Suppose that we choose a target distribution of the form $1_{{\Gamma }%
}(\gamma )\prod\nolimits_{n=1\!}^{P}\eta _{n}(\gamma _{n})$, where each $%
\eta _{n}$ has the Markov property, i.e. $\eta
_{n}^{(s)}(j^{(s)}|j^{(1)},...,j^{(s-1)})=\eta _{n}^{(s)}(j^{(s)}|j^{(s-1)})$%
. Then $\tilde{\eta}_{n}^{(s)}(j^{(s)}|j^{(1)},...,j^{(s-1)})$ = $\tilde{\eta%
}_{n}^{(s)}(j^{(s)}|j^{(s-1)})$, i.e.%
\begin{equation*}
\tilde{\eta}_{n}^{(s)}(j^{(1)},...,j^{(S)})=\dprod\limits_{s=1}^{S}\tilde{%
\eta}_{n}^{(s)}(j^{(s)}|j^{(s-1)}).
\end{equation*}%
and consequently, each normalizing constant reduces to a function of only a
single index as follows 
\begin{eqnarray}
K_{n}^{(1,...,S-1)}(j^{(1)},...,j^{(S-1)}) &=&\sum_{j^{(S)}}\tilde{\eta}%
_{n}^{(S)}(j^{(S)}|j^{(S-1)})\triangleq K_{n}^{(S-1)}(j^{(S-1)}) \\
K_{n}^{(1,...,S-2)}(j^{(1)},...,j^{(S-2)}) &=&\sum_{j^{(S-1)}}\tilde{\eta}%
_{n}^{(S-1)}(j^{(S-1)}|j^{(S-2)})K_{n}^{(S-1)}(j^{(S-1)})\triangleq
K_{n}^{(S-2)}(j^{(S-2)}) \\
&&\vdots  \notag \\
K_{n}^{(1,...,t)}(j^{(1)},...,j^{(t)}) &=&\sum_{j^{(t+1)}}\tilde{\eta}%
_{n}^{(t+1)}(j^{(t+1)}|j^{(t)})K_{n}^{(t+1)}(j^{(t+1)})\triangleq
K_{n}^{(t)}(j^{(t)}) \\
&&\vdots  \notag \\
K_{n}^{(1)}(j^{(1)}) &=&\sum_{j^{(2)}}\tilde{\eta}%
_{n}^{(2)}(j^{(2)}|j^{(1)})K_{n}^{(2)}(j^{(2)})
\end{eqnarray}%
Since we only need to compute $2+M^{(S-1)}$ of the $K_{n}^{(S-1)}(j^{(S-1)})$%
, ...., $2+M^{(t)}$ of the $K_{n}^{(t)}(j^{(t)})$,..., $2+M^{(1)}$ of the $%
K_{n}^{(1)}(j^{(1)})$, the complexity is of order $\mathcal{O}%
(\tsum\nolimits_{s=1}^{S}M^{(s)})$, a drastic reduction from $\mathcal{O}%
(\tprod\nolimits_{s=1}^{S}M^{(s)})$.

To ensure $\tilde{\eta}_{n}^{(s)}(j^{(1)},...,j^{(S)})$ is well-defined on $%
\{-1\}^{S}\uplus \{0:M^{(1)}\}\times ...\times \{0:M^{(S)}\}$, we require
the Markov transition kernel $\tilde{\eta}_{n}^{(s)}(\cdot |j^{(\cdot s-1)})$
to satisfy 
\begin{eqnarray}
\tilde{\eta}_{n}^{(s)}(j^{(s)}|-1) &=&\delta _{-1}[j^{(s)}] \\
\tilde{\eta}_{n}^{(s)}(-1|j^{(s-1)}) &=&\delta _{j^{(s-1)}}[-1]
\end{eqnarray}%
for each $s>1$. In other words, if the chain starts with -1 then each
subsequent state is -1, and if the chain doesn't start with -1 then each
subsequent state cannot take on -1. This implies $K_{n}^{(s)}(-1)=%
\sum_{j^{(s+1)}}\tilde{\eta}_{n}^{(s+1)}(j^{(s+1)}|-1)=\tilde{\eta}%
_{n}^{(s+1)}(-1|-1)=1$, and hence%
\begin{eqnarray}
K_{n}^{(S-1)}(j^{(S-1)}) &=&\left\{ 
\begin{array}{cc}
1, & j^{(S-1)}=-1 \\ 
\sum\limits_{j^{(S)}=0}^{M^{(S)}}\tilde{\eta}_{n}^{(S)}(j^{(S)}|j^{(S-1)}),
& j^{(S-1)}>-1%
\end{array}%
\right. \\
K_{n}^{(s)}(j^{(s)}) &=&\left\{ 
\begin{array}{cc}
1, & j^{(s)}=-1 \\ 
\sum\limits_{j^{(s+1)}=0}^{M^{(s+1)}}\tilde{\eta}%
_{n}^{(s+1)}(j^{(s+1)}|j^{(s)})K_{n}^{(s+1)}(j^{(s+1)}), & j^{(s)}>-1%
\end{array}%
\right.  \label{eq:normalizingconstfor2}
\end{eqnarray}

For $j^{(s)},j^{(s-1)}>-1$, a possible choice of $\eta
_{n}^{(s)}(j^{(s)}|j^{(s-1)})$ is one that is independent of $j^{(s-1)}$,
i.e. $\eta _{n}^{(s)}(j^{(s)})\triangleq \bar{q}^{(\xi ,\ell _{n},s)}\left(
j^{(s)}\right) $, which yields%
\begin{equation}
K_{n}^{(s)}(j^{(s)})=\left\{ 
\begin{array}{cc}
1, & j^{(s)}=-1 \\ 
\Upsilon _{n}^{(s)}\Upsilon _{n}^{(s+1)}...\Upsilon _{n}^{(S-1)} & j^{(s)}>-1%
\end{array}%
\right.  \label{eq:normalizingconstfor1}
\end{equation}%
where 
\begin{eqnarray}
\Upsilon _{n}^{(s)} &\triangleq &\sum_{j^{(s+1)}=0}^{M^{(s+1)}}\tilde{\eta}%
_{n}^{(s+1)}(j^{(s+1)}) \\
\tilde{\eta}_{n}^{(s)}(j^{(s)}) &\triangleq &\eta
_{n}^{(s)}(j^{(s)})\dprod\limits_{i\in \bar{n}}\left(
1-1_{\{1:M^{(s)}\}}(j^{(s)})\delta _{j^{(s)}}[\gamma _{i}^{(s)}]\right)
\end{eqnarray}%
Hence, the conditional distributions for the Gibbs sampler are%
\begin{eqnarray}
\pi _{n}^{(1)}(j^{(1)}) &=&\tilde{\eta}_{n}^{(1)}(j^{(1)})\frac{%
K_{n}^{(1)}(j^{(1)})}{K_{n}} \\
&=&\frac{1}{K_{n}}\left\{ 
\begin{array}{cc}
\tilde{\eta}_{n}^{(1)}(-1), & j^{(1)}=-1 \\ 
\tilde{\eta}_{n}^{(1)}(j^{(1)})\Upsilon _{n}^{(1)}\Upsilon
_{n}^{(2)}...\Upsilon _{n}^{(S-1)}, & j^{(1)}>-1%
\end{array}%
\right. \\
&&\vdots  \notag \\
\pi _{n}^{(t)}(j^{(t)}|j^{(t-1)}) &=&\tilde{\eta}%
_{n}^{(t)}(j^{(t)}|j^{(t-1)})\frac{K_{n}^{(t)}(j^{(t)})}{%
K_{n}^{(t-1)}(j^{(t-1)})} \\
&=&\frac{1}{K_{n}^{(t-1)}(j^{(t-1)})}\left\{ 
\begin{array}{cc}
1, & j^{(t)}=j^{(t-1)}=-1 \\ 
\tilde{\eta}_{n}^{(t)}(j^{(t)})\Upsilon _{n}^{(t)}...\Upsilon _{n}^{(S-1)},
& j^{(t)}>,j^{(t-1)}>-1%
\end{array}%
\right. \\
&&\vdots  \notag \\
\pi _{n}^{(S)}(j^{(S)}|j^{(S-1)}) &=&\tilde{\eta}%
_{n}^{(S)}(j^{(S)}|j^{(S-1)})\frac{1}{K_{n}^{(S-1)}(j^{(S-1)})} \\
&=&\frac{1}{K_{n}^{(S-1)}(j^{(S-1)})}\left\{ 
\begin{array}{cc}
1, & j^{(S)}=j^{(S-1)}=-1 \\ 
\tilde{\eta}_{n}^{(S)}(j^{(S)}), & j^{(S)},j^{(S-1)}>-1%
\end{array}%
\right.
\end{eqnarray}

The alternative target distribution can be interpreted as an approximation
of the original target distribution. When $(j^{(1)},...,j^{(S)})\succeq -1$,

\begin{equation*}
\bar{\psi}_{Z_{+_{\!}}}^{(\xi ,j^{(1)},...,j^{(t)})\!}(\ell _{n\!})=\bar{q}%
^{(\xi ,\ell _{n},1)}\left( j^{(1)}\right) \bar{q}^{(\xi ,\ell
_{n},1,2)}\left( j^{(1)},j^{(2)}\right) ...\bar{q}^{(\xi ,\ell
_{n},1,2,...,t)}\left( j^{(1)},...,j^{(t)}\right)
\end{equation*}%
can be treated as the joint association probability of the measurements $%
z_{j^{(1)}}^{(1)},...,z_{j^{(t)}}^{(t)}$ to track $\ell _{n}$ given the
history $\xi $. Also each $\bar{q}^{(\zeta ,\ell _{n},s)}\left(
j^{(s)}\right) $ can be interpreted as the association probability of
measurements $z_{j^{(s)}}^{(s)}$ to track $\ell _{i\!}$given the history $%
\xi $. In choosing $\eta _{n}^{(s)}(j^{(s)}|j^{(s-1)})=\bar{q}^{(\xi ,\ell
_{n},s)}\left( j^{(s)}\right) $, we are making the simplifying assumption
that the association of measurement $z_{j^{(s)}}^{(s)}$, from sensor $s$, to
track $\ell _{n\!}$ is independent of associations of measurements from
other sensors (to track $\ell _{n}$). Consequently, the joint association
probability of the measurements $z_{j^{(1)}}^{(1)},...,z_{j^{(S)}}^{(S)}$,
to track $\ell _{n}$, is given by the product of the association
probabilities $\bar{q}^{(\xi ,\ell _{n},s)}\left( j^{(s)}\right) $.
Intuitively, if the the product of the association probabilities $\bar{q}%
^{(\xi ,\ell _{n},s)}\left( j^{(s)}\right) $ is high/low, then the joint
association probability of the measurements $%
z_{j^{(1)}}^{(1)},...,z_{j^{(S)}}^{(S)}$ is also high/low.

Remark: If $\bar{\psi}_{Z_{+_{\!}}}^{(\xi ,j^{(1)},...,j^{(t)})\!}(\ell
_{n\!})$, is positive, then the product of the association probabilities $%
\bar{q}^{(\xi ,\ell _{n},s)}\left( j^{(s)}\right) $ is also positive, i.e.
the support of the alternative target distribution contains the support of
the original target distribution. To see this note from (\ref%
{eq:jointassociation}) that $\bar{\psi}_{Z_{+_{\!}}}^{(\xi
,j^{(1)},...,j^{(t)})\!}(\ell _{n\!})$ is actually independent of the order
of the sensors even though we computed it using sensor 1, then sensor 2 and
so on. Further suppose that there exists an $s$ such that $\bar{q}^{(\xi
,\ell _{n},s)}\left( j^{(s)}\right) =0$. Since $\bar{\psi}%
_{Z_{+_{\!}}}^{(\xi ,j^{(1)},...,j^{(t)})\!}(\ell _{n\!})$ is independent of
the order of the sensors, computing it starting from sensor $s$, yields $%
\bar{\psi}_{Z_{+_{\!}}}^{(\xi ,j^{(1)},...,j^{(t)})\!}(\ell _{n\!})=0$.

A more expensive alternative choice is $\eta _{n}^{(s)}(j^{(s)}|j^{(s-1)})=%
\bar{q}^{(\xi ,\ell _{n},s-1,s)}\left( j^{(s-1)},j^{(s)}\right) $. The
additional computation comes from the calculation of normalizing constants (%
\ref{eq:normalizingconstfor2}) rather than (\ref{eq:normalizingconstfor1}).
On the other hand, this choice of $\eta _{n}^{(s)}(j^{(s)}|j^{(s-1)})$
yields a better approximation of $\eta _{n}^{(s)}(j^{(1)},...,j^{(S)})$.
However, since the weights for the components will be corrected after the
Gibbs sampling step, the advantage of a better approximation is not
substantial.

\section{Conclusions}

\label{sec:sum} This paper proposed an efficient implementation of the
Multi-sensor GLMB filter by integrating the prediction and update into one
step along with an efficient algorithm for truncating the GLMB filtering
density based on Gibbs sampling. The resulting algorithm is an on-line
multi-sensor multi-object tracker with linear complexity in the number of
measurements of each sensor and quadratic in the number of hypothesized
tracks. This implementation is also applicable to approximations such as the
labeled multi-Bernoulli (LMB) filter since this filter requires a special
case of the GLMB prediction and a full GLMB update to be performed \cite%
{ReuterLMB14}.

%\IEEEtriggeratref{41} % The "triggered" command can be changed if desired:
%%\IEEEtriggercmd{\enlargethispage{-5in}}
%
%% references section
%\bibliographystyle{IEEEtran}
%\bibliography{IEEEabrv,D:/Research/Tex/References/GLMB_ref}

\providecommand{\url}[1]{#1} \csname url@samestyle\endcsname%
\providecommand{\newblock}{\relax} \providecommand{\bibinfo}[2]{#2} %
\providecommand{\BIBentrySTDinterwordspacing}{\spaceskip=0pt\relax} %
\providecommand{\BIBentryALTinterwordstretchfactor}{4} 
\providecommand{\BIBentryALTinterwordspacing}{\spaceskip=\fontdimen2\font plus
\BIBentryALTinterwordstretchfactor\fontdimen3\font minus
  \fontdimen4\font\relax} 
\providecommand{\BIBforeignlanguage}[2]{{\expandafter\ifx\csname l@#1\endcsname\relax
\typeout{** WARNING: IEEEtran.bst: No hyphenation pattern has been}\typeout{** loaded for the language `#1'. Using the pattern for}\typeout{** the default language instead.}\else
\language=\csname l@#1\endcsname
\fi
#2}} \providecommand{\BIBdecl}{\relax} \BIBdecl

\end{document}